# Plasmonic Graphene Waveguides: A Literature Review

**Mohammad Bagher Heydari [1], Mohammad Hashem Vadjed Samiei[1,*]**

[1] Electrical Engineering Department, Iran University of Science and Technology (IUST), Tehran, Iran
[*]Corresponding author: mh_samiei@ iust.ac.ir

**ABSTRACT:**
A large research articles based on graphene have been published in various areas since 2007 due to its unique properties. Among these designed structures, graphene-based plasmonic waveguides are one of interesting subjects in communications. In this paper, a historical review of plasmonic graphene waveguides is presented. The graphene based waveguides can be divided into two main categorizes: 1- rectangular and 2- cylindrical waveguides. In both categorizes, graphene can be only electrically biased or magnetically and electrically biased simultaneously where the conductivity of graphene becomes a tensor in second case. Therefore, we have four main categorizes for plasmonic graphene waveguides which are studied and discussed more precisely in this paper.

**Key words:** Graphene, Rectangular waveguide, cylindrical waveguide, electric bias, magnetic bias

## 1. INTRODUCTION

The emerging two-dimensional (2D) materials like molybdenum disulfide ( $MoS_2$ ) and tungsten diselenide ( $WSe_2$ ) have generated immense interest for semiconductor and nanotechnology due to their fascinating properties [1-5]. Among these 2D materials, carbon nanotubes exhibit attractive features which are widely used in chemistry and physics [6, 7]. Graphene is 2D sheet which offers a number of fundamentally superior properties that has the potential to enable a new generation of technologies [8-10].

The usage of graphite started 6000 years ago[11]. The research on graphene has grown slowly in late 20$^{th}$ century but AB initio calculations illustrated that a single graphene layer is unstable [12]. Andre Geim and Konstantin Novoselov first isolated single layer samples from graphite and Nobel prize in physics 2010 was awarded to them [13, 14].

Graphene is one of the allotropes of carbon, which is planar monolayer of carbon atoms that forms a honey-comb lattice with carbon-carbon bond length of 0.142 nm [15]. It exhibits a large wide of interesting properties [16-21]. One of them is its highly unusual nature of charge carries which behave as Dirac fermions [22]. This feature has a great effect on energy spectrum of landau levels produced in presence of magnetic field [23, 24]. The Hall conductivity is observed at zero energy level. Also, the Hall Effect is distinctly different than conventional Hall Effect which is quantized by a half integer [25, 26].

Graphene absorbs 2.3% of incident light and the absorption can be tuned by varying the Fermi level through the electrical gating [27, 28]. This property is widely used in designing transparent electrodes in solar cells [29-31]. Graphene has exceptional thermal conductivity (5000 $w/m^{-1}$ $K^{-1}$) that is utilized in fabricating and designing the electronic components [11]. Graphene transport characteristics and electrical conductivity can be tuned by either electrostatic or magneto static gating or via chemical doping, which this conductivity leads to design the various photonic and electronic devices [32-39]. Among the large designed devices in papers using this feature, linear Plasmonic waveguides have great attention due to their applications in communications.

This paper represents a literature review for linear Plasmonic graphene waveguides. In section 2, rectangular and planar graphene waveguides are reviewed in detail. These waveguides have two categories: graphene waveguides where graphene has electrical bias and graphene waveguides with magnetic and electric bias. Then, in section 3, cylindrical waveguides with similar categories are introduced and reviewed. Finally, section 4 concludes the paper.

## 2. RECTANGULAR GRAPHENE WAVEGUIDES

In this section, various planar waveguides have been introduced and discussed in detail. These graphene waveguides can be divided to two main categories: 1- Graphene waveguides with electrically biased graphene. 2- Graphene waveguides with anisotropic graphene sheet (magnetically biased graphene). In Fig. 1, various rectangular graphene waveguides have been illustrated. This section gives a historical review to rectangular graphene waveguides.

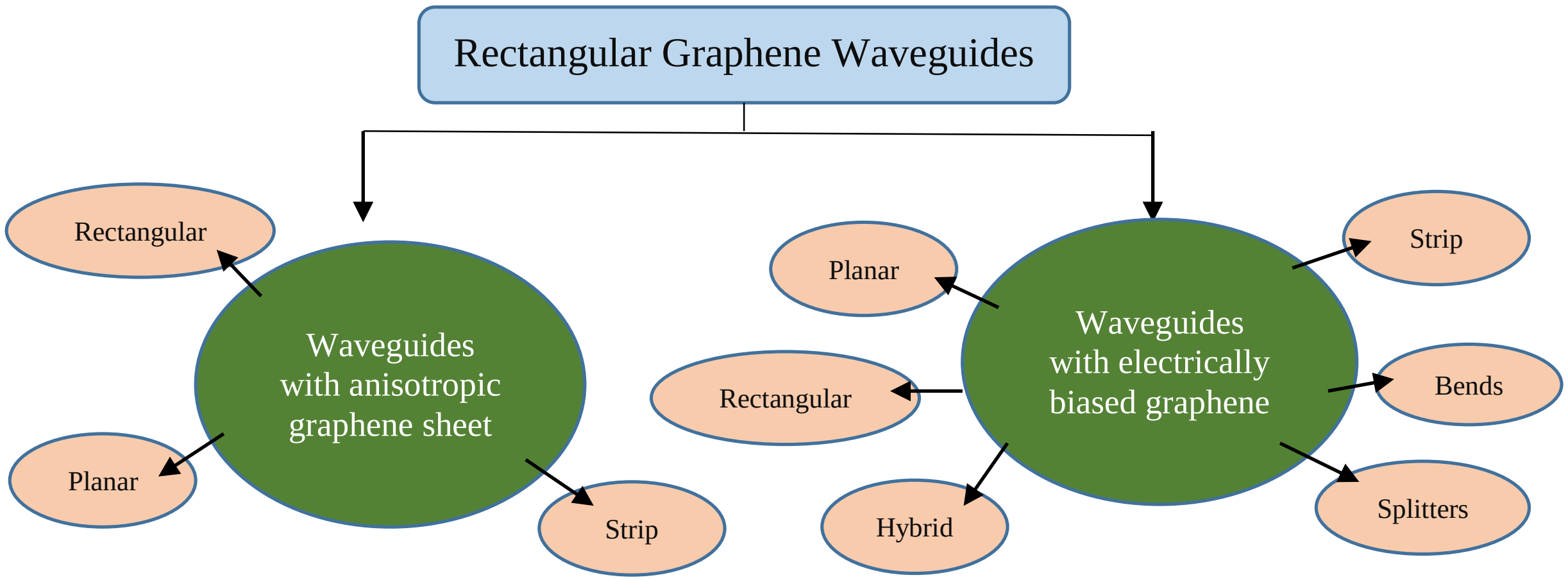

Fig. 1. Rectangular Graphene Waveguides

## 2.1 Rectangular Graphene Waveguides with Electrically Biased Graphene

Many researchers have begun to work on graphene waveguides since 2007. There are many articles that considered the various graphene waveguide structures. In this section, rectangular graphene waveguides have been discussed in detail, where graphene sheet is electrically biased in all of these structures. As seen in Fig.1, these waveguides can be categorized to various structures such planar, strip, etc. Fig. 2 represents number of publications for each kind of various waveguides. It is obvious that planar graphene waveguides have maximum publications among other types of waveguides. All of these structures use Kubo equation for conductivity of graphene [40]:

$$\sigma\left(\omega,\mu_c,\Gamma,T\right)=\frac{je^2\left(\omega-j2\Gamma\right)}{\pi\hbar^2}\left[\begin{array}{l}\dfrac{1}{\left(\omega-j2\Gamma\right)^2}\displaystyle\int_0^\infty \varepsilon\left(\frac{\partial f_d\left(\varepsilon\right)}{\partial\varepsilon}-\frac{\partial f_d\left(-\varepsilon\right)}{\partial\varepsilon}\right)d\varepsilon\\ -\displaystyle\int_0^\infty \frac{f_d\left(\varepsilon\right)-f_d\left(-\varepsilon\right)}{\left(\omega-j2\Gamma\right)^2-4\left(\varepsilon/\hbar\right)^2}\left(\frac{\partial f_d\left(\varepsilon\right)}{\partial\varepsilon}-\frac{\partial f_d\left(-\varepsilon\right)}{\partial\varepsilon}\right)d\varepsilon\end{array}\right] \quad (1)$$

Where

$$f_d\left(\varepsilon\right)=\frac{1}{1+\exp\left(\dfrac{\varepsilon-\mu_c}{K_BT}\right)} \quad (2)$$

In above equations, $e$ is the charge of an electron, $\hbar$ is the reduced Planck's constant, $K_B$ is Boltzmann's constant, $T$ is temperature, $\omega$ is radian frequency and is chemical potential. By substituting (2) into (1) and integrating [11]:

$$\sigma\left(\omega,\mu_c,\Gamma,T\right)=\sigma_{\text{inter}}(\omega)+\sigma_{\text{intra}}(\omega)=\frac{-je^2}{4\pi\hbar}Ln\left[\frac{2\left|\mu_c\right|-(\omega-\mathrm{j}2\Gamma)\hbar}{2\left|\mu_c\right|+(\omega-\mathrm{j}2\Gamma)\hbar}\right]+\frac{-je^2K_BT}{\pi\hbar^2(\omega-\mathrm{j}2\Gamma)}\left[\frac{\mu_c}{K_BT}+2Ln\left(1+e^{-\mu_c/K_BT}\right)\right] \quad (3)$$

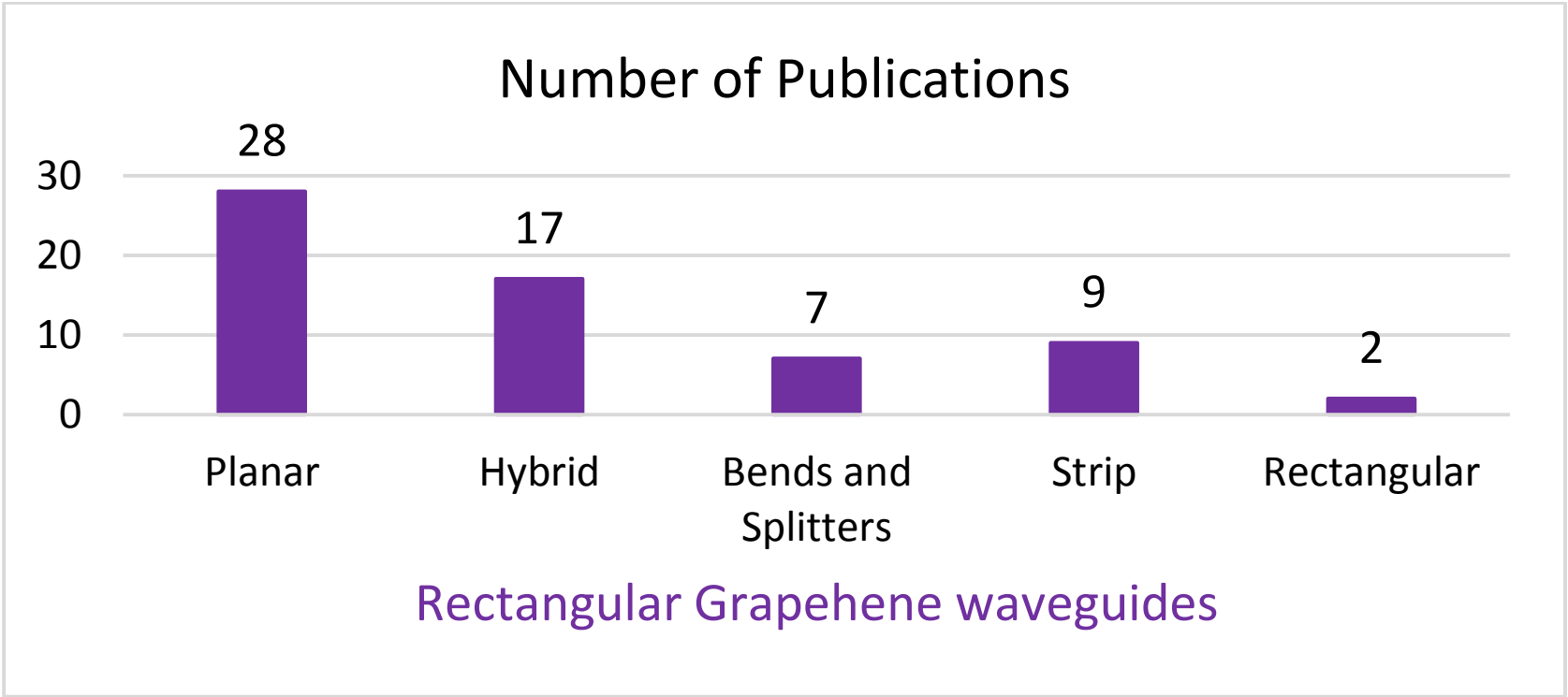

Fig. 2. Number of publications on rectangular graphene waveguides with electrically biased graphene from 2007 to 2017

### *2.1.1 Planar waveguides*

Nowadays, planar THz waveguides are used in photonic integrated technology due to their unique properties. They are easy to design, simplify the theoretical analysis and have easy fabrication process. Fig.3 represents number of publications for planar graphene waveguide articles in recent years.

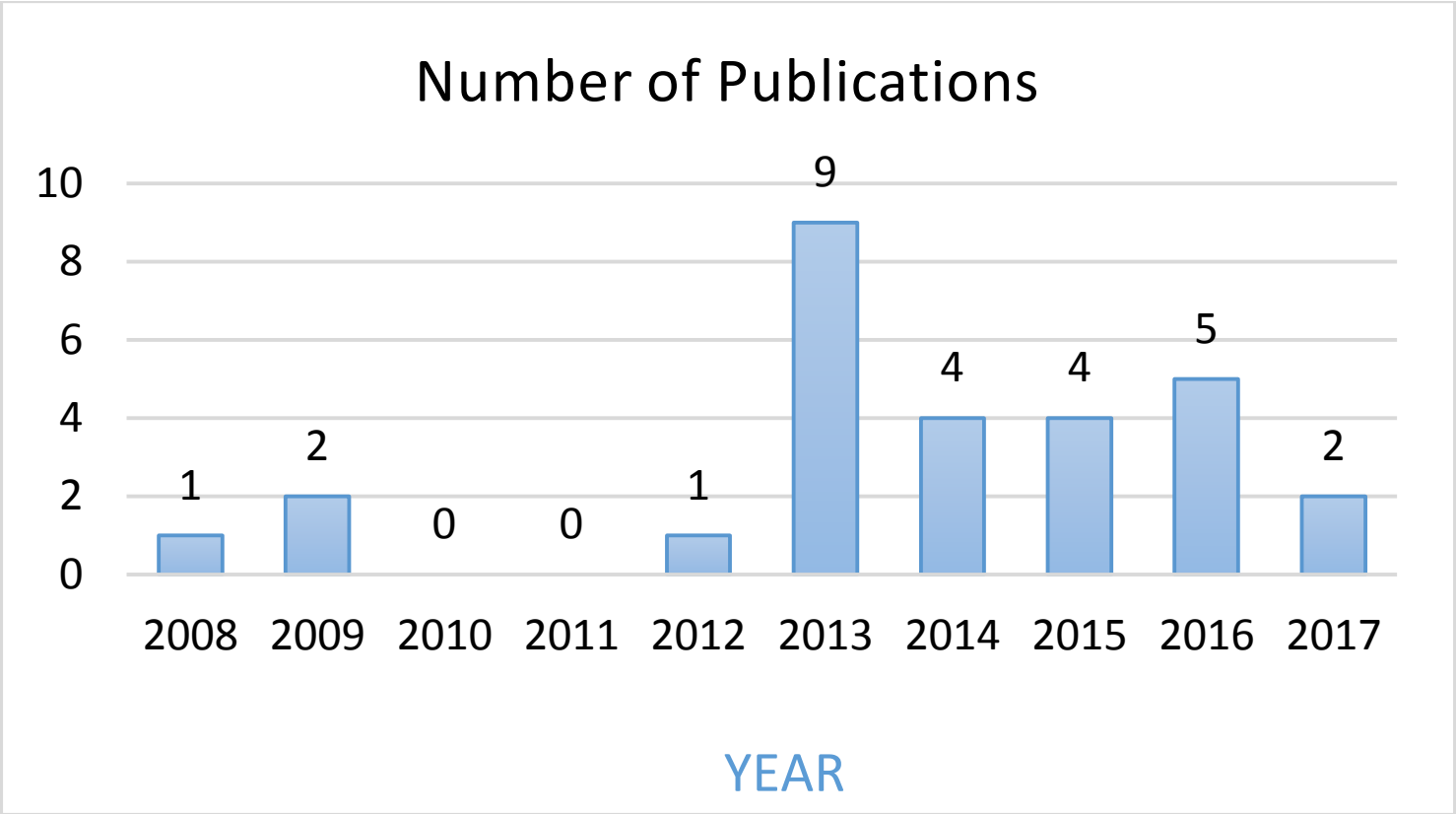

Fig. 3. Number of publications on planar graphene waveguide from 2008 to 2017

A first study on graphene parallel plate waveguide was done by George W. Hanson in 2008 [41]. In this article, Maxwell's equations were solved to derive a model for quasi-transverse electromagnetic mode (quasi-TEM) propagating in the parallel plate waveguide formed by graphene layers. A more detail study for graphene layer waveguide was reported in [42]. A famous dispersion relation for TM modes propagating on the graphene sheet sandwiched between two dielectric layers have been derived in this research which is used in many articles now [42]:

$$\frac{\varepsilon_{r1}}{\sqrt{q^2-\frac{\varepsilon_{r1}\omega^2}{c^2}}}+\frac{\varepsilon_{r2}}{\sqrt{q^2-\frac{\varepsilon_{r2}\omega^2}{c^2}}}=-\frac{\sigma(\omega,q)i}{\omega\varepsilon_0} \tag{4}$$

Where $\varepsilon_{r1},\varepsilon_{r2}$ relative permittivity of dielectrics, $q$ is wave number and conductance of the 2D graphene layer are is $\sigma(\omega,q)$.

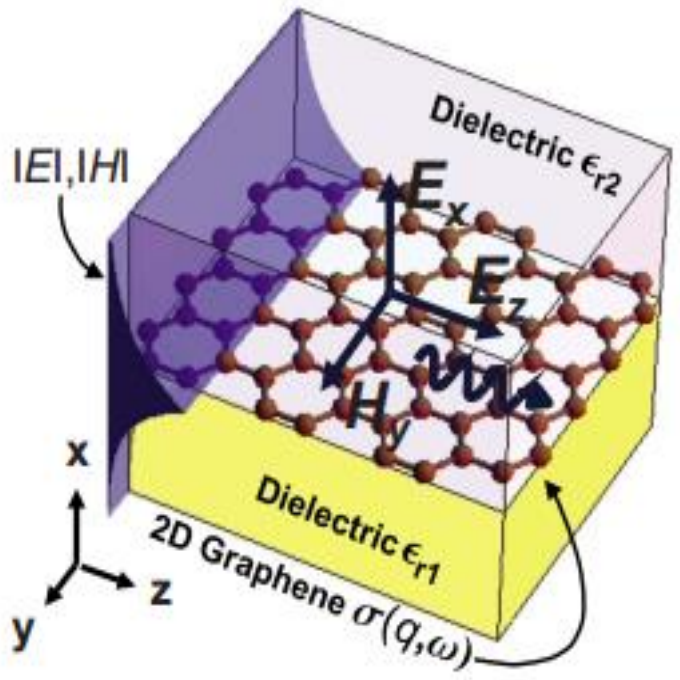

Fig. 4. The graphene sheet sandwiched between two dielectric layers with relative permittivity of $\varepsilon_{r1},\varepsilon_{r2}$ [42]

Fan-Ming Zhang et al. have investigated planar graphene waveguides based on two different dispersion relations (classical motion and Klein tunneling) in [43]. Researchers have used Dirac-type equation for symmetric quantum well of graphene waveguides and the wave function of guided modes as function of distance have been plotted and discussed in this paper [43]. Giampiero Lovat studied the dominant mode propagation in graphene nano-waveguides which a mono layer graphene is deposited on top of the grounded dielectric [44]. He derived dispersion relations by an exact approach based on transverse equivalent network (TEN) for different graphene nano waveguides. The TEN for the structure of fig. 5(a) is shown.

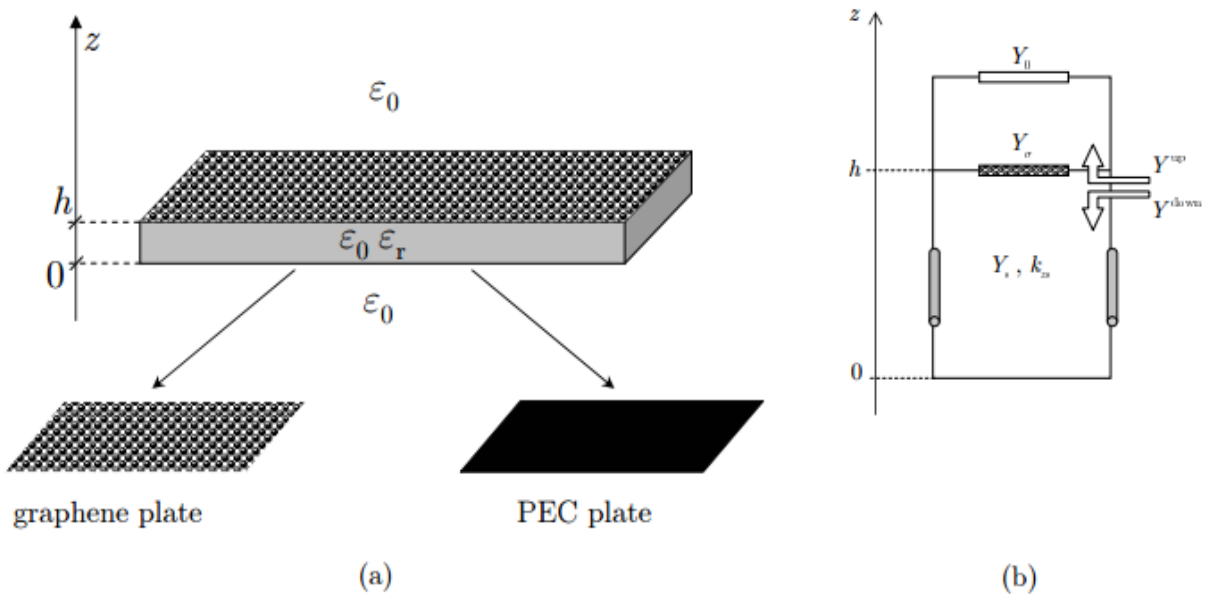

Fig. 5. (a) Graphene nano-waveguides composed by a mono layer graphene located on top of the grounded dielectric with thickness of $h$ and permittivity $\varepsilon_r$ , (b) Transverse-equivalent network for the waveguide [44]

A numerous research articles have been published for graphene planar waveguides in 2013, as seen in fig. 3. This matter confirmed that scientists have paid attention to planar structures due to its simplicity and easy to design. In [45], the dispersion characteristics of graphene Plasmonic waveguides have been considered. The transverse magnetic mode (TM) can be propagated when the imaginary part of graphene conductivity (see relation (1)) is positive. The authors have used this property to design plasmonic graphene waveguide which supports TM plasmons in middle area where imaginary part of graphene conductivity is positive over there [45]. Hybridization of graphene-metal for planar waveguides was proposed in [46]. The waveguide focused in this work was similar to the waveguide of [44] but this research had an accurate look to modal properties. The authors have achieved a good mode confinement with low loss which can be utilized in many integrated THz devices such modulators. Also, it is shown that by narrow slits of a proper periodicity in metal, the SPP mode can be existed under normal incident beam [46]. The article [47] focuses on a graphene sheet deposited on silicon carbide (SiC) and depicts that the electromagnetic local density of states (EM-LDOS) which peak positions of these states depend on chemical potential and therefore EM-LDOS can be controlled by $\mu_c$ .The authors in [48] have investigated the structure constructed of graphene sheet sandwiched between two dielectrics, similar to [42].

Xiao Yong HE and Rui LI have done a comparison between transverse magnetic (TM) and electric (TE) surface plasmon modes (SPPs) for air-graphene-$SiO_2$-Si structure [49]. This waveguide consists of graphene layer located on the $SiO_2$/Si layers, as represented in fig. 6. The authors derived the dispersion relations for TM and TE SPPs by starting from Maxwell's equations and applying boundary conditions. The effective mode indices and propagation losses for TE and TM modes have been exhibited for air-graphene-$SiO_2$-Si structure for various thickness of the $SiO_2$ layer ( $d = 50, 100, 200, 300\ nm$ ) in the article [49]. The temperature is 3 $K$ and the Fermi level is assumed 0.2 $eV$. It has been concluded that TM mode is bound (lossy) mode while the TE is lossy mode and effective index of the graphene supported TM mode increases as the operation frequency increases. The authors deduce that TM mode has better confinement that conventional metal-based structures [49].

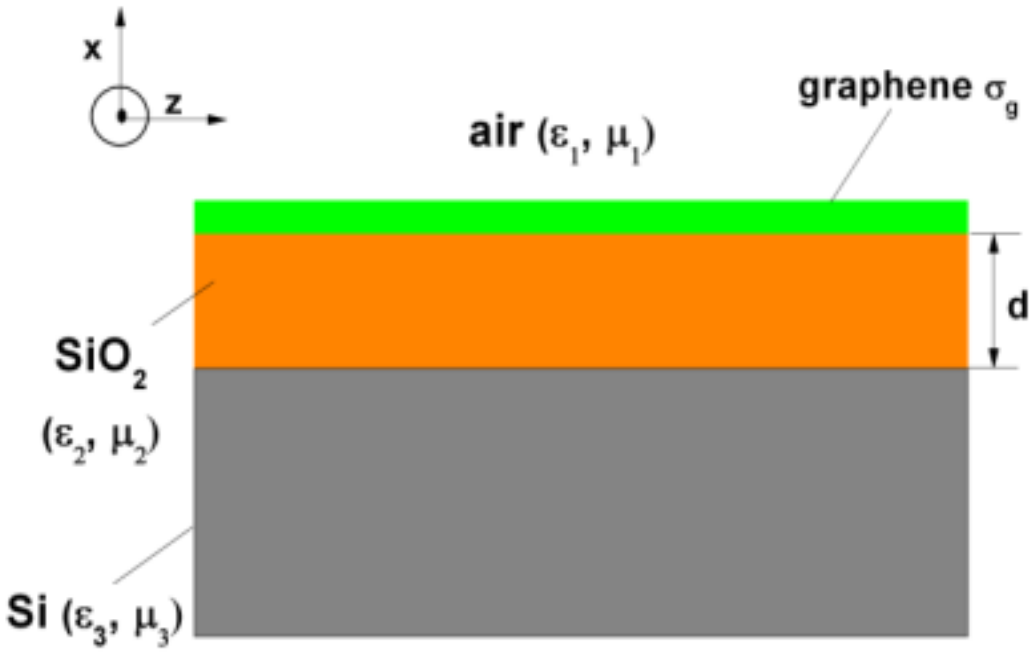

Fig. 6. The schematic of air-graphene-$SiO_2$-Si structure [49]

A dielectric-graphene-dielectric waveguide based on multi-layer metamaterials have been presented in [50]. This periodic waveguide is simulated in COMSOL software and results reveal that the device can be utilized in nonlinear optics and bio sensing because it has high field confinement and refractive indices. Also, the modal properties of the structure can be controlled by graphene conductivity [50]. Similar to [44, 49], plasmons in waveguides formed by two graphene layers has been studied for both TE and TM-polarized waveguide modes [51]. Diego Correas-Serrano et al. have proposed the circuit model for graphene single and parallel plate waveguides which can be accurately determine the propagation [52]. The authors have considered influence of the spatial dispersion on modal characteristics such as losses and confinement. The numerical results confirmed that spatial distribution must be taken to account for graphene-based plasmonics structures [52]. Optimization of double layer graphene plasmonic waveguides has been studied in [53]. To minimize the attenuation of the SPP propagation, an optimum frequency and an optimum spacing between two graphene layers has been achieved by considering both intrinsic landau damping and extrinsic scattering [53]. Field confinement of stack graphene slab waveguide is done in [54]. In this paper, a numerical coding is solved for finding propagation constant for TE field. To determine the influence of the number of graphene layers on dispersion and localization of guided modes, a more detailed study is reported by D. A. Smirnova et al [55]. In this research, the graphene waveguide composed of N-layers of graphene where are separated with a dielectric material with thickness *d* and permittivity $\varepsilon$ . The dispersion results indicates that by using multi-layer graphene structures, the degree of localization of plasmon modes can be controlled [55].
Xiao-Qiang Gu et al. introduced a novel parallel plate graphene waveguide by applying a ferroelectric layer [56], as demonstrated in fig. 7. Multi-layer graphene sheet has been placed on the grounded ferroelectric layer with thickness of *d*. vertically light with energy of $\hbar\Omega$ illuminates the whole structure. In this work, the guided THz waves characteristics of this structure are analyzed and dispersion diagrams are depicted [56]. The authors believe that the designed device can be used as THz waveguide especially for where the waveguide with low and tunable attenuation is needed.

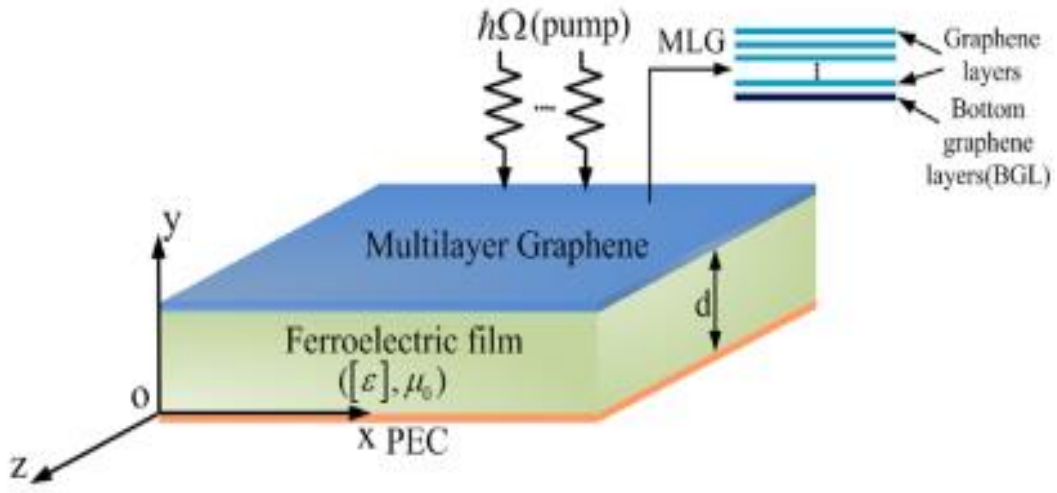

Fig. 7. The schematic of parallel plate graphene waveguide with optical pumping [56]

In 2015, there are some articles which are similar to previous works but in different context. For instance, Walter Fuscaldo et al. have considered the modal properties of a graphene based planar waveguide in THz frequency range [57]. The work is similar to articles [44, 46] which the structure composed by a graphene sheet on the grounded dielectric slab. For both guided and radiative (leaky) regimes, the dispersion relation is derived and the structure is simulated at $f = 1$ *THz*. In the rest of the paper, the effect of losses has been investigated [57]. The authors of [58] have studied the TM and TE modes of graphene SPP in three-layered structure, similar to article [41]. Chengzhi Qin et al. investigated the multi-layer graphene waveguide [59], similar to [55]. In this work, normalized TM field distribution based on FDTD computation has been achieved by using the transfer matrix method (TMM) for TM polarized SPPs propagating in *z* direction. In [60], based on graphene parallel plate waveguide (which formed by two graphene layers spaced a distance apart), a THz power divider with tunable power division ratio has been introduced. The authors have displayed the simulation results of S-parameters for 5 THz with different power ratio. The results indicate that the contours of the electric length of graphene parallel plate waveguide varies with the operation frequency and chemical potentials which means that the proposed device is tunable [60].
In recent years, scientists tend to fabricate the Graphene applied structures and do not be satisfied only to analytical and simulation results. The [61] is one the articles which reports an experimental verification of optical modes of multi-mode slab waveguides based on graphene at the wavelengths of 632.8 nm and 1535 nm. A comparison of three optical model of graphene (interface model, isotropic model and anisotropic model) with experimental results indicate that interface and anisotropic model are more accurate that the isotropic model for TE and TM modes [61]. The structure of fig. 6 has been studied more precisely in [62], which has similar context to [49]. In this paper, the authors investigate modal cut-off properties for the waveguide by FEM. It is concluded that reduction of the width of graphene decreases the effective index of waveguide rapidly and single mode operation can be realized by reducing W [62].
Similar to articles [41, 51, 52, 56] that have studied the three-layered graphene waveguides, a three-layered graphene waveguide where a gyro electric slab sandwiched between two graphene layers with different conductivities (denoted as $\sigma_1, \sigma_2$) has been investigated in [63], as demonstrated in fig. 8. The half spaces is constructed by two dielectric materials with permittivity of $\varepsilon_1, \varepsilon_2$, as illustrated in the figure.

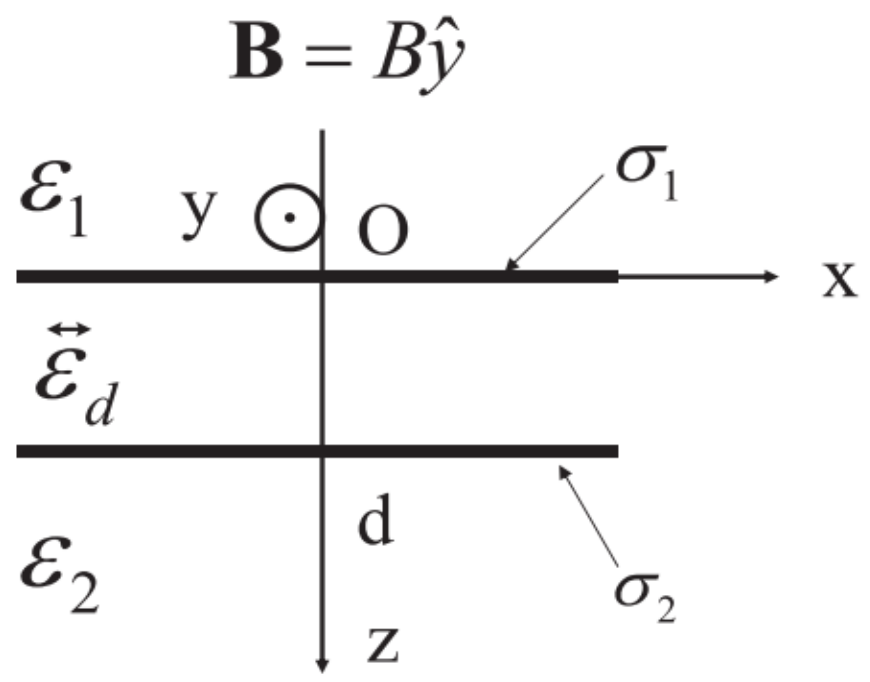

Fig. 8. A gyro electric slab sandwiched between two graphene layers with different conductivities [63]

The authors consider dispersion and propagation properties of SPPs for various external magnetic fields for this waveguide [63]. As mentioned in this work, the non-reciprocal properties of SPPs caused by the external magnetic field is a method for adjusting the propagation properties where is used for designing tunable plasmonic devices [63]. The article [64] is one of other works which studies the leaky and bound modes of graphene SPPs in three-layered graphene waveguides. By numerically solved the dispersion equation for TM polarized SPPs, the results illustrate that the chemical potential of graphene can influence the radiation efficiency and frequency which is important in THz regimes such as designing the THz plasmonic antennas [64]. A graphene sandwiched between two dielectric layers is one of the interest structures that researchers tend to study new aspects of it, as reported in [42, 47, 48, 52]. Again, a new article investigates the mentioned structure by solving the Schrodinger equation to consider the behavior of quasi-TM or quasi-TE modes [65].

The plasmonic graphene waveguides can be utilized to design useful photonic devices. Zeshan Chang et al. have proposed a graphene-embedded waveguide as a filter [66]. In the mentioned structure, a graphene film has been located in the central axis of symmetric slab waveguide. The authors have achieved a mode extinction ratio higher than 20 dB against the $TE_0$ mode in experimental results [66]. In the article done by Vahid Foroughi Nezhad et al, a graphene parallel plate waveguide has been used to design convertors and optical diodes [67]. The reciprocal convectors are proposed based on spatial modulation of graphene conductivity and optical diodes are composed by a mode convector and a coupler [67].

### *2.1.2 Hybrid waveguides*

Hybrid plasmonic waveguides provide large confinement of light at a lower loss compared to many previously plasmonic waveguides. Hybridization in plasmonic graphene waveguides can be divided to main categorizes: 1- Metal-graphene hybrid waveguides, 2- Hybrid Graphene plasmonic waveguide with dielectric waveguide or with other shape waveguides. We will review these kinds of hybridization with historical aspect in this section. Fig. 9 represents number of publications for planar graphene waveguide articles in recent years.

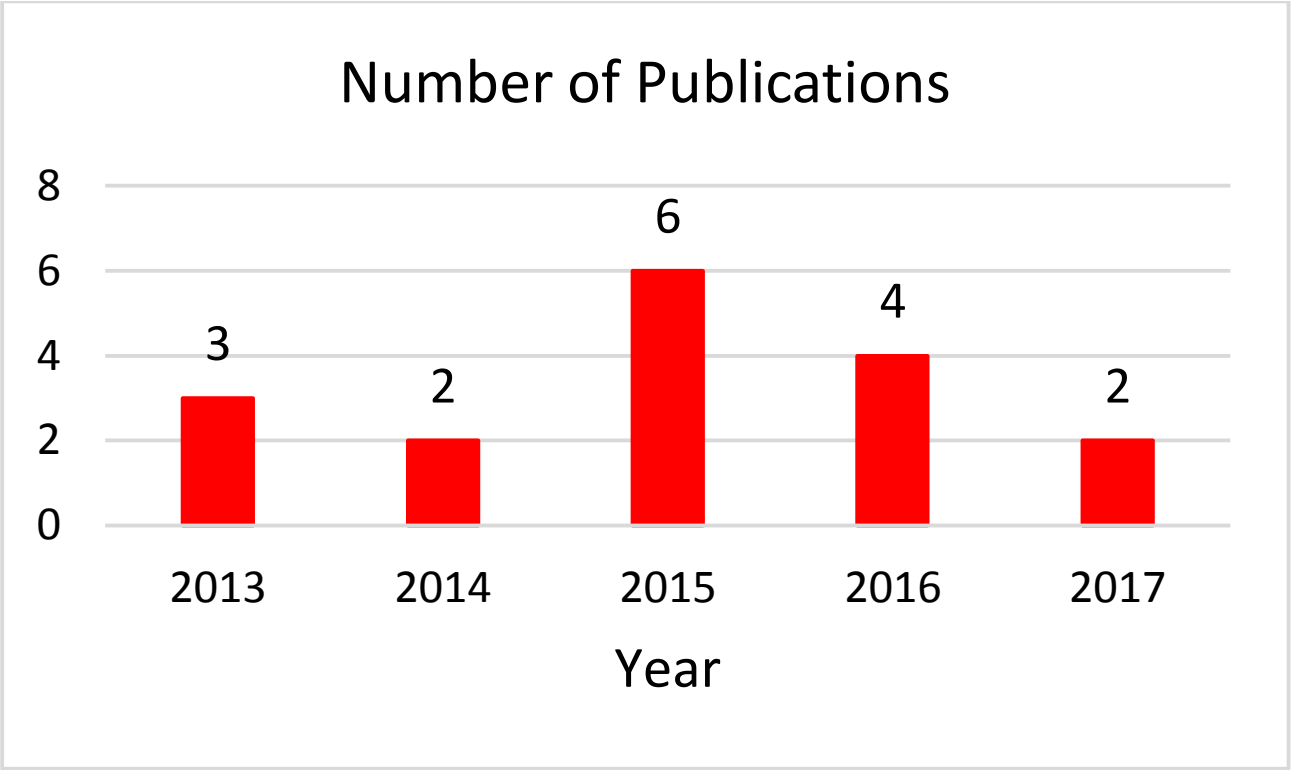

Fig. 9. Number of publications on hybrid graphene waveguide from 2013 to 2017

First hybrid graphene waveguides were published in 2013 [68-70]. Two hybrid multi-layered tunable waveguides based on graphene are introduced in [68], as represented in fig. 10. First structure is composed by Si–$SiO_2$–graphene–dielectric-stripe–metal layers while the second waveguide is consists of Si–$SiO_2$–graphene–dielectric–graphene–

$SiO_2$–Si layers. $SiO_2$ has been used in both structures to apply the gate voltage to bias the graphene sheet. TMM is applied to analyze these structures to obtain mode area. The authors concluded from the results that the designed hybrid waveguides have smaller the propagation loss than the conventional MDM structures which can be used in many plasmonic devices such switches and polarizers [68].

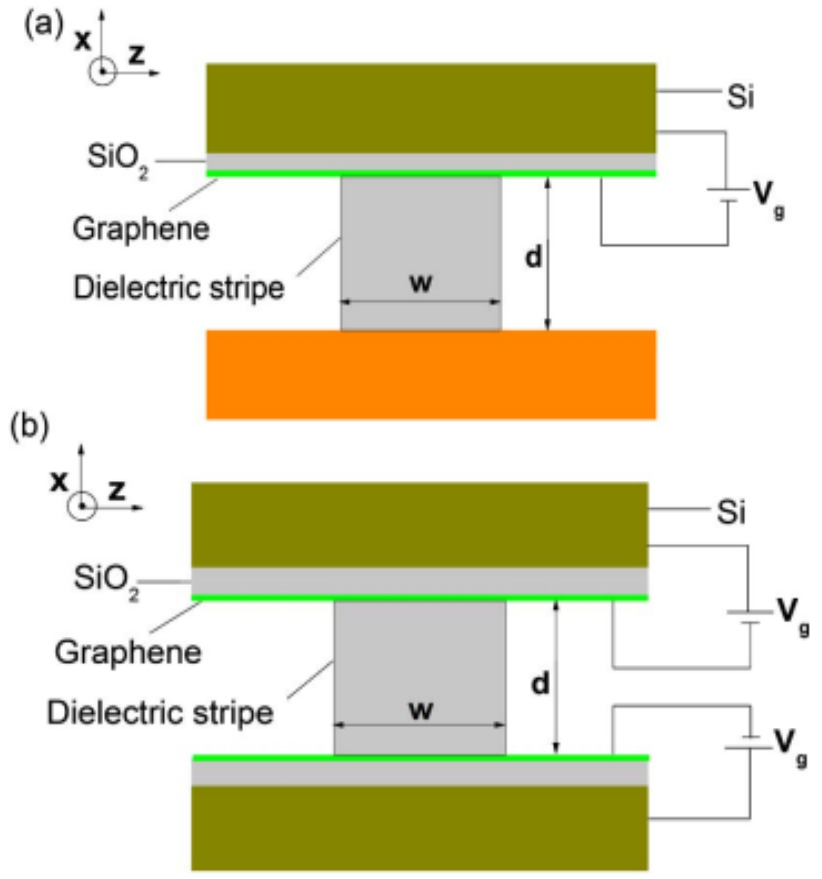

Fig. 10. The hybrid plasmonic graphene waveguides: (a) Si–$SiO_2$–graphene–dielectric-stripe–metal, (b) Si–$SiO_2$–graphene–dielectric–graphene–$SiO_2$–Si [68]

Yu Sun and his coworkers have presented a novel hybrid plasmonic waveguide where KCL (*n=1.46*) is deposited on the GaAs ridge (*n=3.3*) with height *h* and width of *w* and a graphene sheet is placed at top of the KCL [69, 70], as seen in fig. 11. $|E(x, y)|$ distributions of the fundamental $TM_{00}$ and $TM_{01}$ modes are exhibited for various gap sizes. It is obvious that the power is concentrated in gap region and the hybrid waveguide can be operated single mode. The authors claim that their structure has stronger field distribution and lower propagation loss that the convectional graphene ribbon waveguides [69, 70]. A novel graphene based hybrid plasmonic waveguide has been reported in [71], where high density polyethylene (HDPE) and GaAs are high index (*n=3.6*) and low index (*n=1.54*) dielectric layers in the structure, respectively. Multi-layer graphene has thickness of $t = 2.5\ nm$ and the structure is simulated in COMSOL. The authors achieved small mode area of $32.6\ \mu m^2$ and propagation length of $127\ \mu m$ at 3 THz [71]. Also, the crosstalk of two integrated waveguides are discussed more precisely in this work.

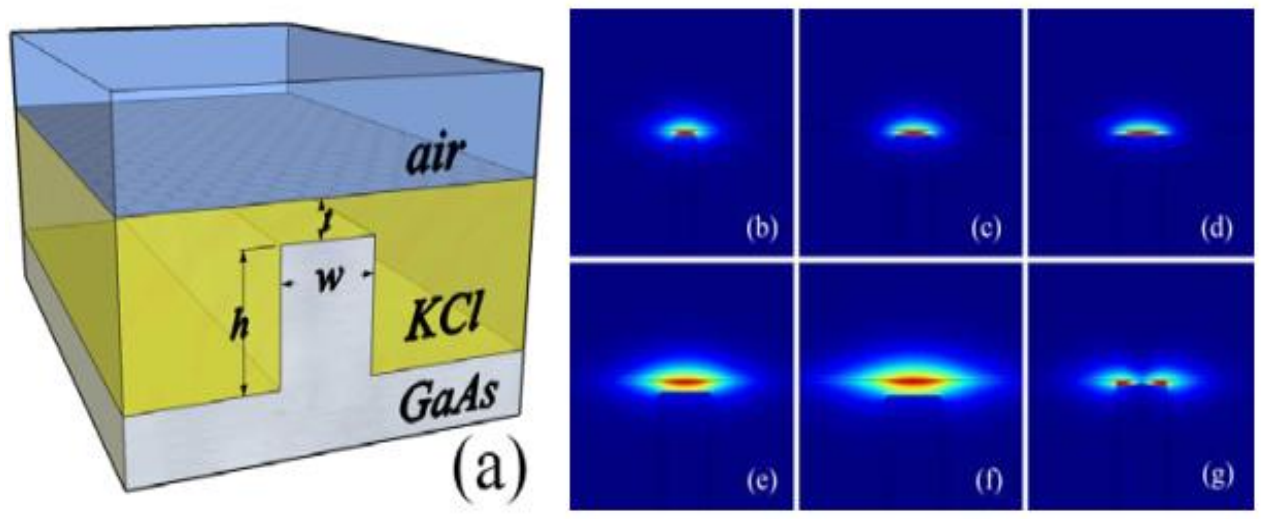

Fig. 11. (a) Hybrid plasmonic ridge waveguide where KCL is deposited on the GaAs ridge and a graphene sheet is placed at top of the KCL; (b-f) $|E(x, y)|$ distributions of the fundamental $TM_{00}$ mode for: (b)-(d) *w=40, 70, 100 nm* (*t=7 nm*), respectively, (e)-(f) *t= 20, 25 nm* (*w=100 nm*), respectively, (g) $|E(x, y)|$ distributions of $TM_{01}$ mode for: *t=7 nm* and *w= 100 nm* [70]

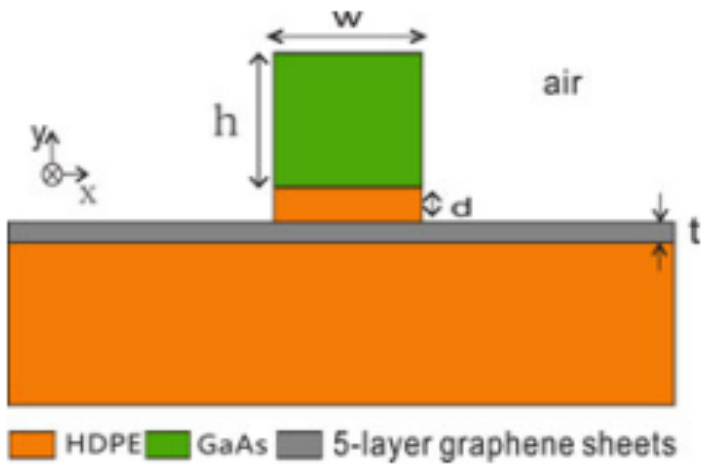

Fig. 12. The schematic of hybrid plasmonic waveguide where high density polyethylene (HDPE) and GaAs are high index (*n=3.6*) and low index (*n=1.54*) dielectric layers in the structure. The geometrical dimensions are $w = 20\ \mu m$, $t = 2.5 nm$ and excitation wavelength is $100\ \mu m$ [71]

I-Tan Lin and Jia-Ming Liu have been designed a hybrid graphene-metal waveguide where the core layer with thickness of $d_2+d_3$, width of $w$ and permittivity of $\varepsilon_{core}$ has been located above and below the graphene sheet and the dielectric with permittivity of $\varepsilon_{cladding}$ has been placed between two metal slabs [72]. A semi analytical approach is considered in this paper to achieve the dispersion relation for the hybrid waveguide. The authors have expressed that by hybridization of graphene-metal, the plasmonic attenuation length and the attenuation length normalized to the plasmonic wavelength have been increased 20% and 97%, respectively [72].

As seen in fig.9, there is an increment in publications of graphene hybrid plasmonic waveguides in 2015. Researchers utilize hybrid graphene structures to design novel plasmonic devices. For instance, Myunghwan Kim et al. have presented an optical modulator with modulation depth of 100% based on graphene-metal hybrid plasmonic waveguide at 1550 nm [73]. The proposed modulator is tunable because the conductivity of graphene can be adjusted to give the best performance [73]. The article [74] focuses on nonlinearity effect based on hybrid graphene-metal plasmonic waveguide. The authors concluded that large dispersion values of the proposed structure (in the order of $10^4$ ps/km-nm) can be applied to design short ranged femtosecond solitons [74]. Novel graphene-silicon hybrid waveguide is introduced by Qiang Jin et al. to study the nonlinearity enhancement in near infrared band (the wavelength of 1580 nm) [75]. This device utilizes $MoO_3$ layer in order to adjust the chemical potential of graphene to achieve good conversion efficiency (-27.3 dB) in this band [75].

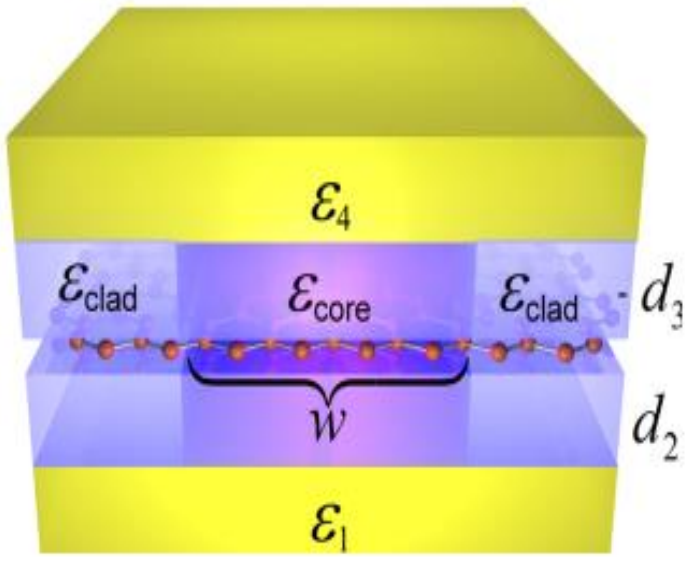

Fig. 13. Hybrid graphene-metal waveguide where $\varepsilon_1, \varepsilon_2$ represents the metal slabs and core layer has been deposited below and above the graphene. The cladding dielectric is placed between two metal slabs [72]

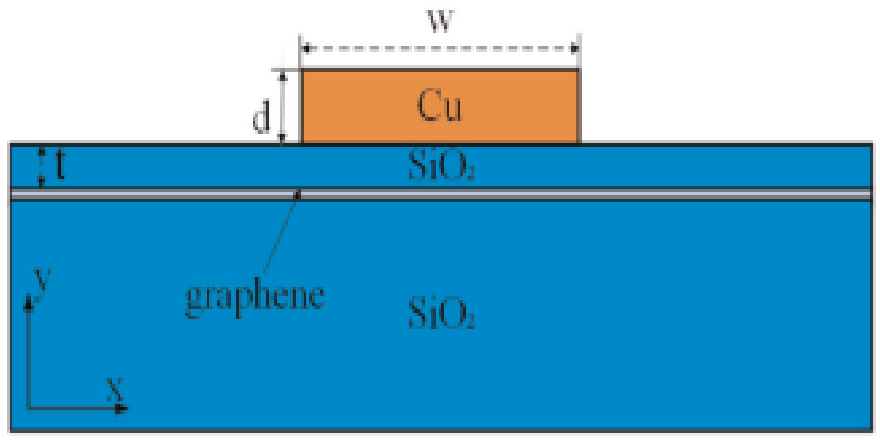

Fig. 14. Hybrid graphene-metal plasmonic waveguide introduced in [76].

Another metal-graphene hybrid plasmonic waveguide is displayed in fig. 14 where the copper ($\sigma = 4\times10^7\ S/m$) strip with width $w$ and height $d$ is deposited on top of $SiO_2$ ($n=3.9$) as the spacer material and graphene is placed between two $SiO_2$ layers [76]. The authors have studied the modal properties of waveguide of TM mode and dependence of the mode effective index on various parameters such chemical potential [76]. To investigate the device packing density, a cross talk between adjacent waveguides has been done in the article [76]. W. Xu et al. have been reported an easy-to-design dielectric loaded graphene plasmon waveguide (DLGPW) where a dielectric strip has been placed on a graphene sheet [77]. The effective index method (EIM) is applied to analyze the waveguide at mid-infrared region and the modal properties like cut-off wavelength of higher order modes are studied in the article [77]. A strip assisted hybrid plasmonic waveguide at $f=3$ $THz$ have been designed in [78]. The proposed hybrid waveguide is simulated in COMSOL and the results show good propagation length of about $820\ \mu m$, power confinement ratio of 0.65 and normalized mode area of 0.03[78]. The authors conclude that this structure improves normalized mode area and power ratio in comparison of conventional hybrid plasmonic waveguide [78]. Fig. 16 demonstrates symmetric long range SPP hybrid waveguides based on graphene layers which has introduced by Jian-Ping Liu et al [79]. The authors apply coupled mode perturbation theory to analyze the structure where detailed explanations are existed in the article [79].By using FDTD simulations, field distributions and modal properties such as effective mode index and propagation length can be achieved. The authors reported propagation length of 10 $\mu m$ and mode area of $10^{-7} A_0$ and finally concluded that their proposed hybrid waveguide can be utilized to design compact devices for integrated photonics [79].

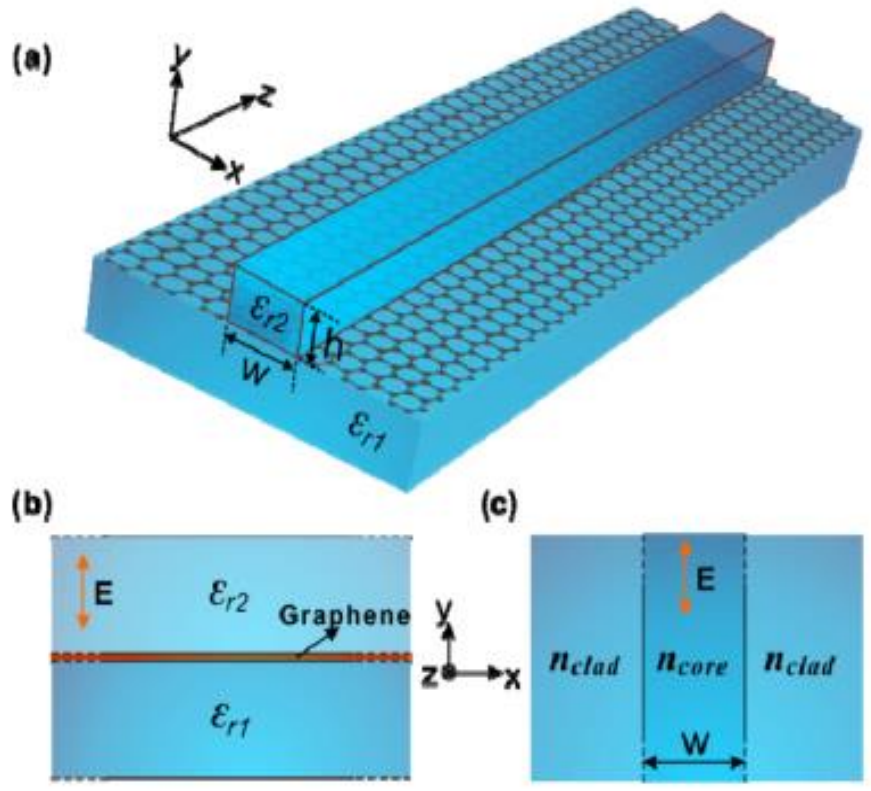

Fig. 15. (a) dielectric loaded graphene plasmon waveguide, (b,c) the steps of effective index method (EIM) [77]

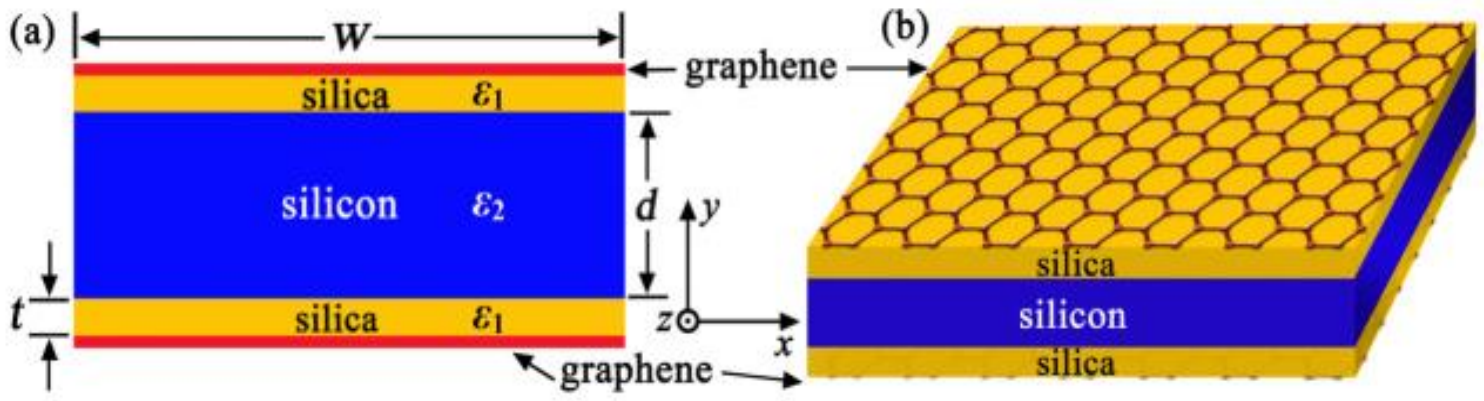

Fig. 16. (a) Cross section and (b) 3D view of hybrid graphene based plasmonic waveguide [79]

In recent years, the researchers tend to present applied devices based on hybrid plasmonic waveguides or study other aspects of these plasmonic waveguides. For instance, Jun Guo and his co-workers focuses on tunable fano resonances of graphene hybrid waveguide in [80]. They consider the tunability of fano resonances with graphene conductivity to suggest other scientists for designing novel sensors and switches based on this fano resonances. In [81], the authors have reported the applied structures based on hybrid graphene-silicon waveguides. They first introduce the silicon-graphene switches based on Mach-Zehnder Interferometer (MMI) where these MMI couplers can be used as power splitters and combiner [81]. Then, logic gates (AND/NAND) are designed based on MMI and studied in detail.

Yan Liu and his coworkers presented a new hybrid graphene waveguide with efficient interference suppression in 41st International Conference on 2016 Sep 25 *IEEE* [82] and developed their research in [83]. The mentioned waveguide has been composed by double dielectric ridges (Si) where are separated by a low index dielectric ($SiO_2$) and graphene sheet is located at the middle of the structure, as shown in following figure [83]. It is obvious that the SSPs are confined in the spacer region, as exhibited in fig 17 (b). One can see from fig 17 (c,d) that all the higher order modes ($m$ > 0) have cut off as wavelength increases except the fundamental mode ($m$ = 0). The author discuss about crosstalk between two adjacent hybrid waveguides and concluded that the low crosstalk compared with conventional plasmonic waveguide has been obtained [83].Another novel hybrid plasmonic waveguide based on hybridization of metal and graphene is presented by Hua Lu et al. in [84]. In this research, a metal-dielectric-metal plasmonic waveguide with a sandwiched graphene mono layer is theoretically analyzed and then by using the graphene at epsilon-near-zero point, a sharp peak in the propagation attenuation can be seen. Graphene can be tuned by the voltage bias in this waveguide to achieve an active modulation with an extinction ratio of 15.8 db [84].

### *2.1.3 Curved waveguides, Bends and splitters*

Bends and power splitters have important role in plasmonic circuits. A first study on bends and splitters in graphene waveguides has been done in [85, 86]. In these papers, first a 90 sharp graphene bend is investigated and studied more precisely. As illustrated in fig. 18 (a), the normalized transmission shows that there is no additional loss at longer wavelength and the wavelength that transmission coefficient reduces below 99% is 12 and 14 micrometer for *d=20 nm* and *d=30 nm*, respectively [85]. Then, the authors propose a graphene based splitter for *d= 20, 30 nm*. Figure 18 (b) exhibits the transmission coefficients as function of wavelength. This figure indicates that the transmission saturates at the longer wavelengths and the transmitted power is equally divided between two branches of splitter due to symmetry [85].

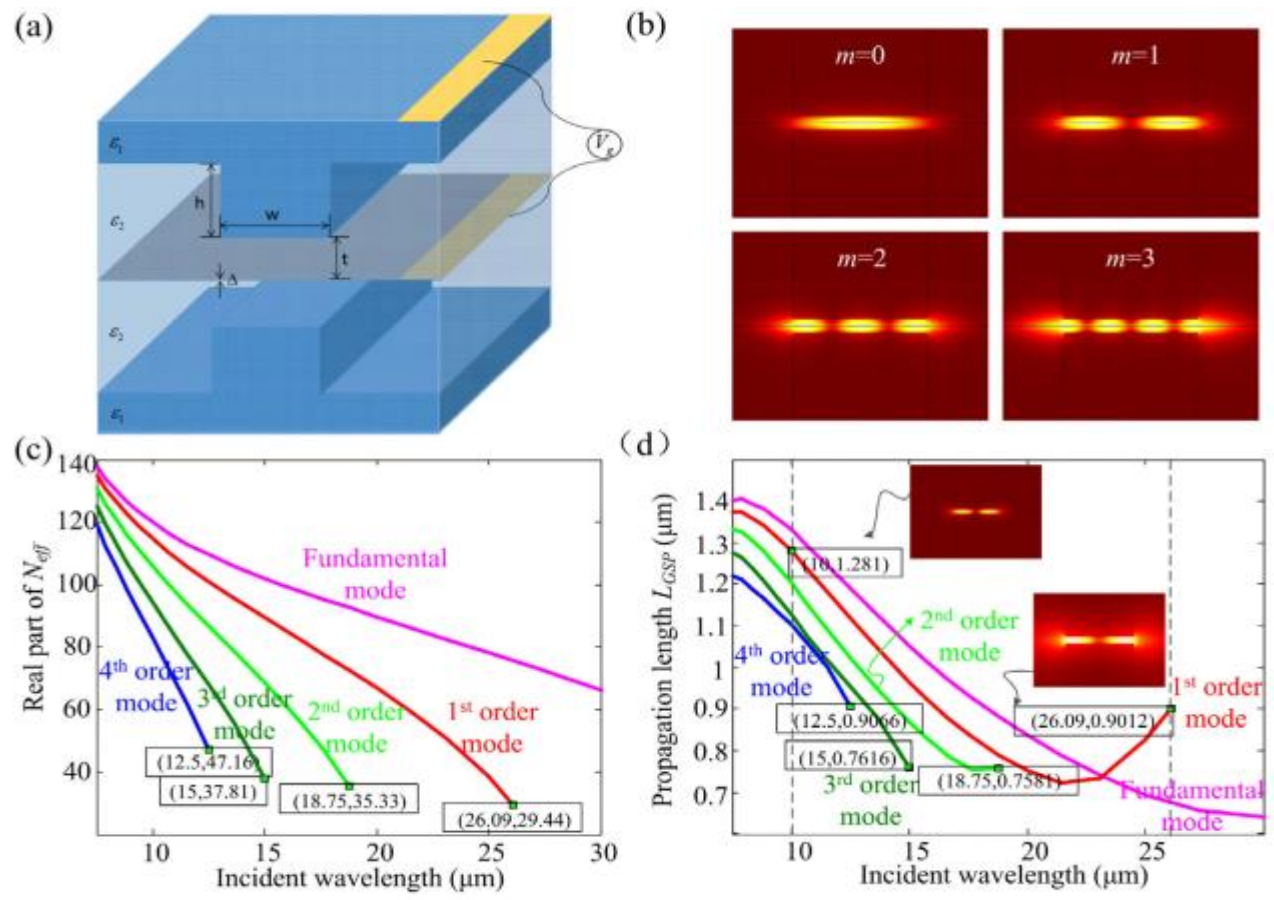

Fig. 17. (a) The hybrid graphene plasmonic waveguide, (b) Mode patterns, (c) The real part of the effective index (*Re* (*Neff*)) and (d) Propagation length $L_{GSP}$ for the first four SPP modes at the wavelength of 10 $\mu$m and $E_f$ = 0.2 eV. The geometrical parameters are $W_{ridge}$ = 200 nm and $2t$ = 20 nm [83]

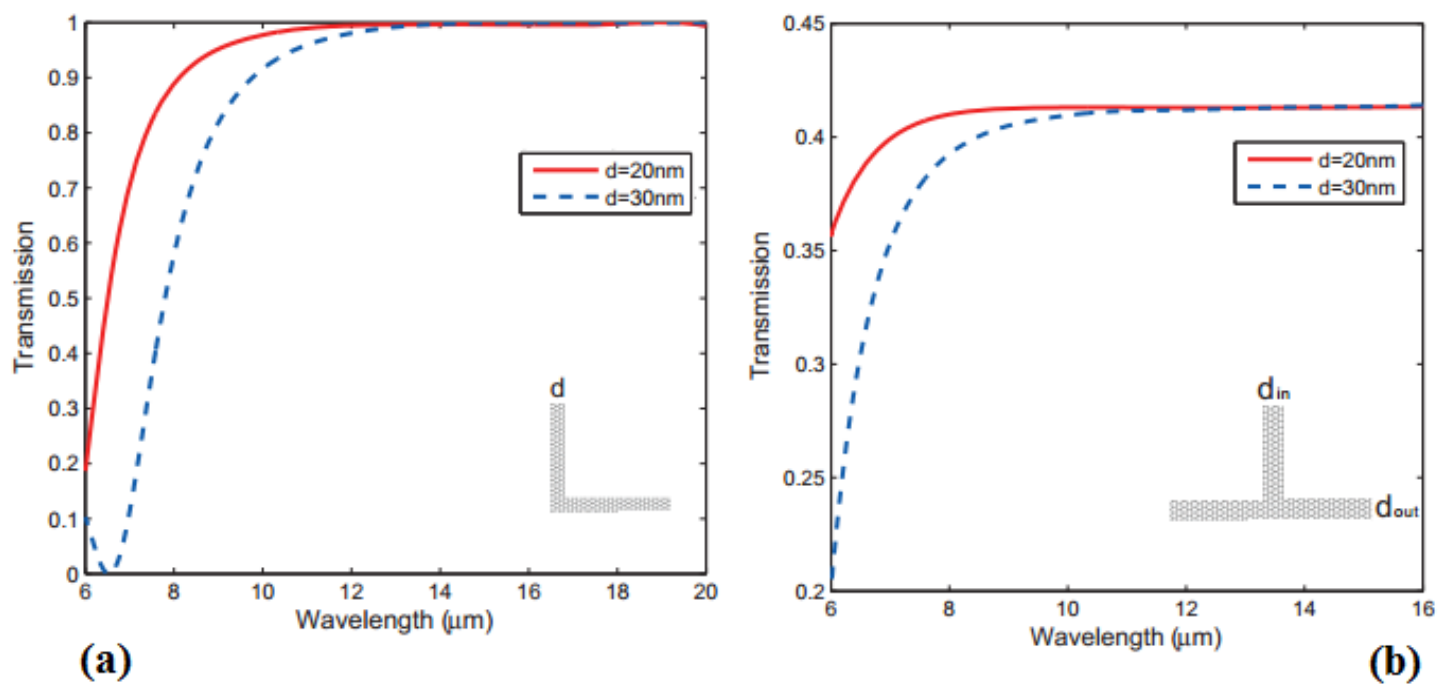

Fig. 18. Transmission as a function of wavelength for $d=d_{in}=d_{out}=$ *20, 30 nm*: (a) Bend, (b) Splitter [85]

Nowadays, the flexible plasmonic devices have been considered due to their unique applications. For instance, Wei Bing Lu and his co-workers have introduced novel bend and splitter flexible plasmonic waveguides which are on the curve surfaces [87]. This paper focuses on flexible transmission plasmonic using graphene and discusses the simulated results more precisely. The magnetic field $H_y$ distributions of various bending and splitter waveguides at *f= 160 THz* are illustrated in fig. 19. The authors have indicated that using graphene in curve flexible plasmonic devices can be enhance the confinement of SPPs in comparison with metals [87]. In fig 19. (a), a side view of Y-shaped splitter on a sinusoidal curve with function $z(x) = 12\sin(\pi x/60)$ nm in the region $0 \le x \le 300$ *nm* .The coupling of edge modes in graphene bend waveguides are studied in [88]. They authors of this paper have shown that modal fields of two edge modes shift in opposite directions. This new phenomena occurs due to bending effect [88]. Ting-Hui Xiao and his coworkers have investigated graphene SPPs on curved substrates [89]. They first present an analytical model for cylindrical substrate with graphene coating which is curved substrate with a fixed radius and then study propagating SPPs and modal properties such propagation length for U-shaped, S-shaped, G-shaped and $90^0$ graphene bend [89].

Applying bends and splitters to design novel plasmonic devices are one the interesting subjects in graphene plasmonics. For instance, Hu Jian-rong et al. present s-shaped waveguide attenuator based on graphene that can carry THz waves [90]. The attenuator consists of a doped silicon substrate, a $SiO_2$ buffer layer and graphene layer which has two sections with different chemical potentials, as exhibited in fig. 20 [90]. The tuned extinction ratio from 1.28 to 39.42 db by change chemical potential in range 0.03-0.05 ev is reported in this work [90]. The authors deduced that the proposed attenuator has several advantages like simplicity, tunability and low insertion loss [90]. In another article in 2016, a sinusoidally curved graphene structure is introduced for obtaining plasmonically induced transparency (PIT) [91], similar to [87]. To analyze the mentioned structure, coupled mode theory has been applied that the reader can refer to the article [91].

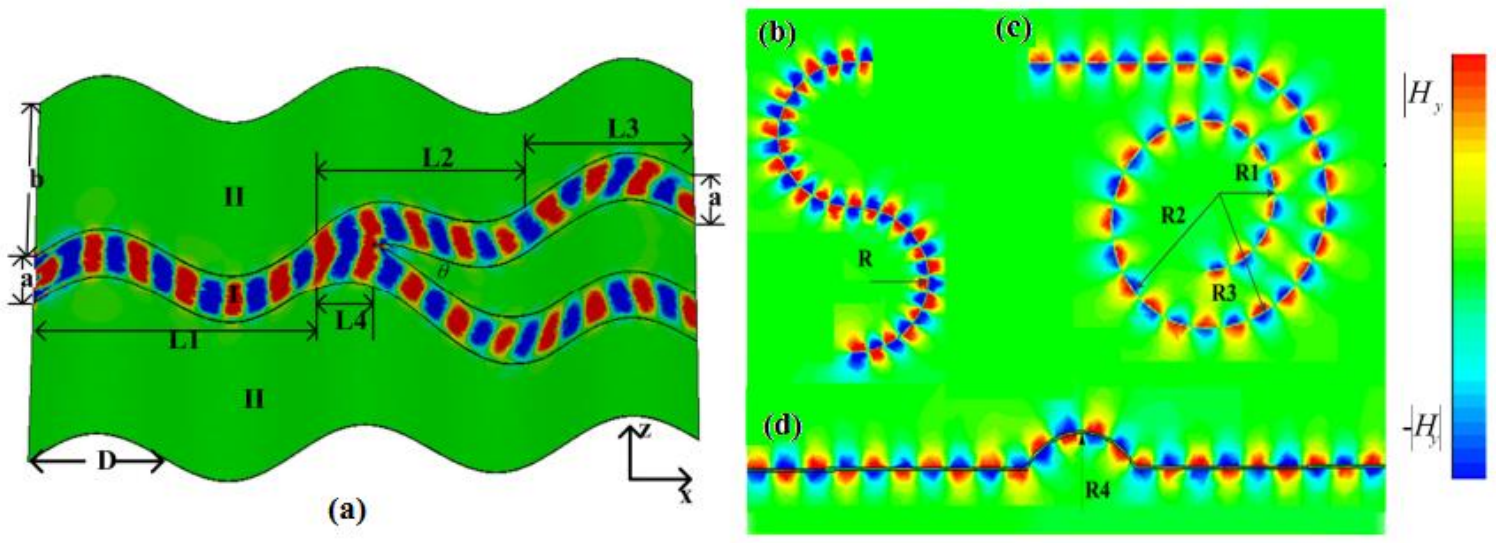

Fig. 19. The magnetic field $H_y$ distributions on: (a) the Y-shape flexible waveguide, (b)-(d) the curved graphene surfaces at *f= 160 THz* [87].

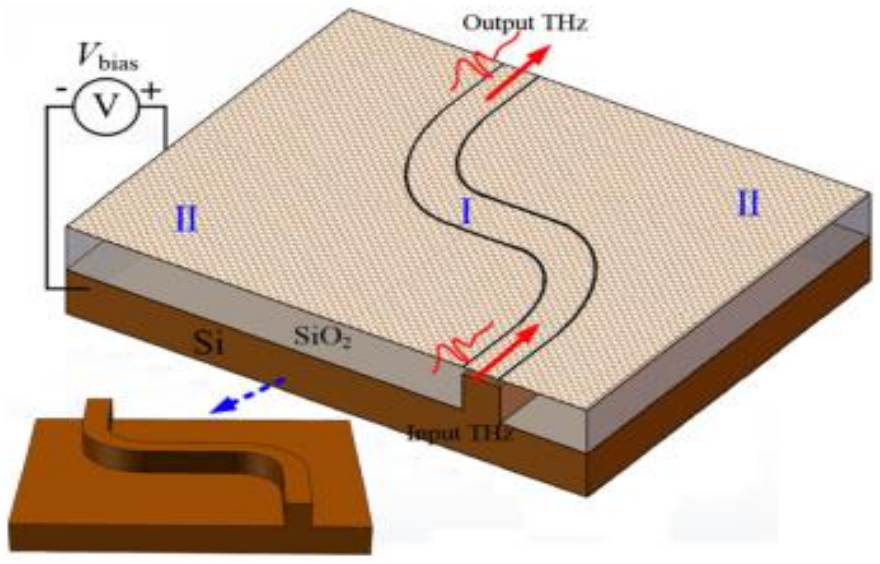

Fig. 20. Graphene based attenuator [90]

### *2.1.4 Strip waveguides*

Graphene nano ribbon (GNR) is graphene sheet with finite width in nano dimension. A numerous articles has been published in GNRs for various applications. However, we focus on GNR as graphene strip waveguides in this section. Therefore, this section intends to summarize all graphene strips with finite width which can be introduced as nano-ribbon, micro ribbon, etc. First important study on individual and paired GNRs as plasmonic waveguides is done in 2011 by Johan Christensen, et al [92]. The authors in this paper have investigated the modal properties of the individual and paired GNRs, as illustrated in fig. 21[92]. The more detailed explanations about modal properties of GNRs is given in the context of this paper [92]. The article [93] is one of the main articles in Graphene micro ribbon which numerous articles have been referred. This paper considers the waveguide and edge modes propagated in graphene micro ribbon in detail for a graphene strip with width of *w* located between two dielectrics, as demonstrated in fig. 22 [93]. It should be noted that number of SPPs propagating in graphene ribbon can be increased when the frequency or the ribbon width increases [93].

J. S. Gómez-Díaz et al. present a different aspect from graphene strip waveguide [94]. They study the graphene strips propagating non resonant plasmons to design nano devices such tunable leaky-wave antenna where the width-modulated graphene is deposited on the dielectric, as shown in fig. 23 [94]. The poly silicon layer has been used for applying the DC bias voltage of graphene. This periodic modulated strip acts as leaky-wave antenna which the leaky-wave mode radiates to air. The fig. 23 (e). confirms the beam scanning behavior of the proposed antenna [94].

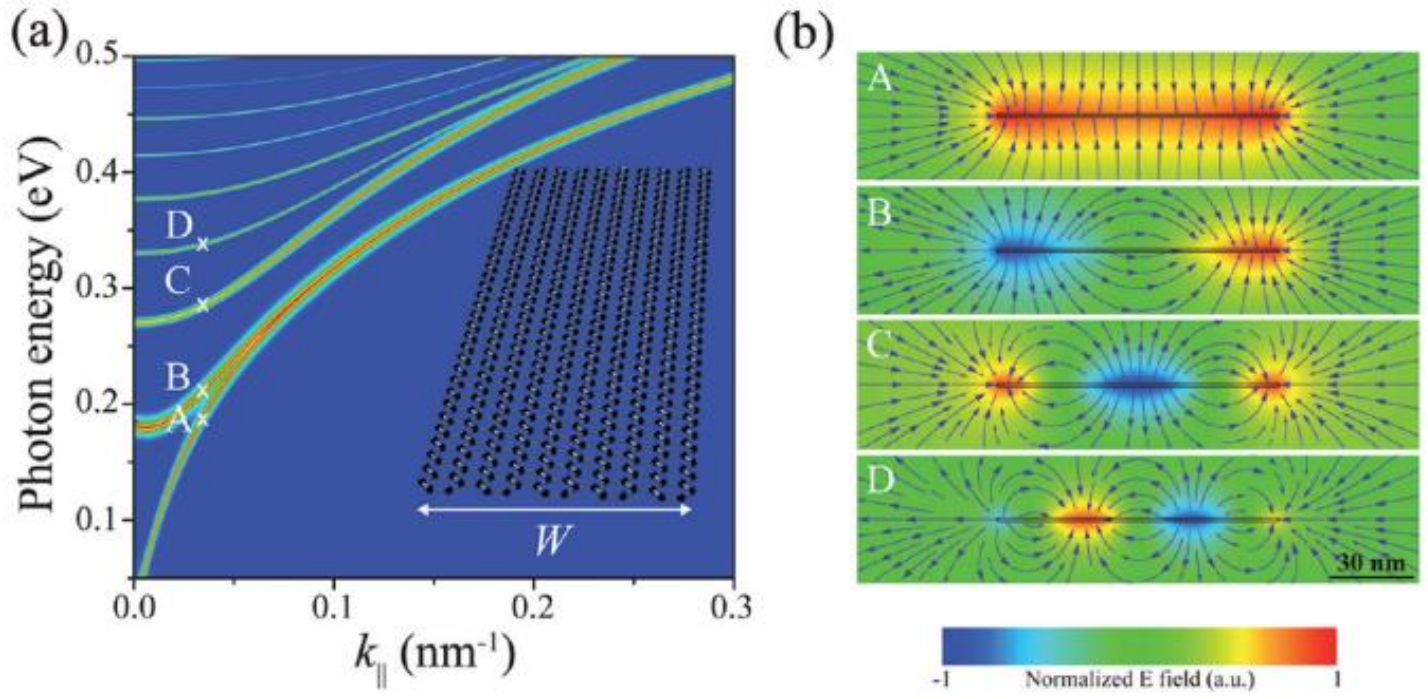

Fig. 21. (a) Dispersion diagram of single GNR for *w=100 nm*,
(b) Real part of electric field amplitude for modes labeled as A-D in fig.21 (a) [92].

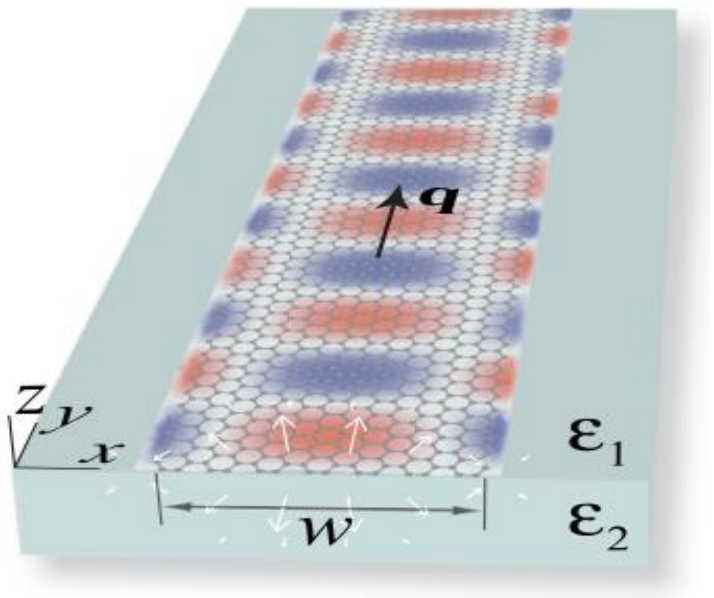

Fig. 22. The graphene micro ribbon has been sandwiched between two dielectrics [93].

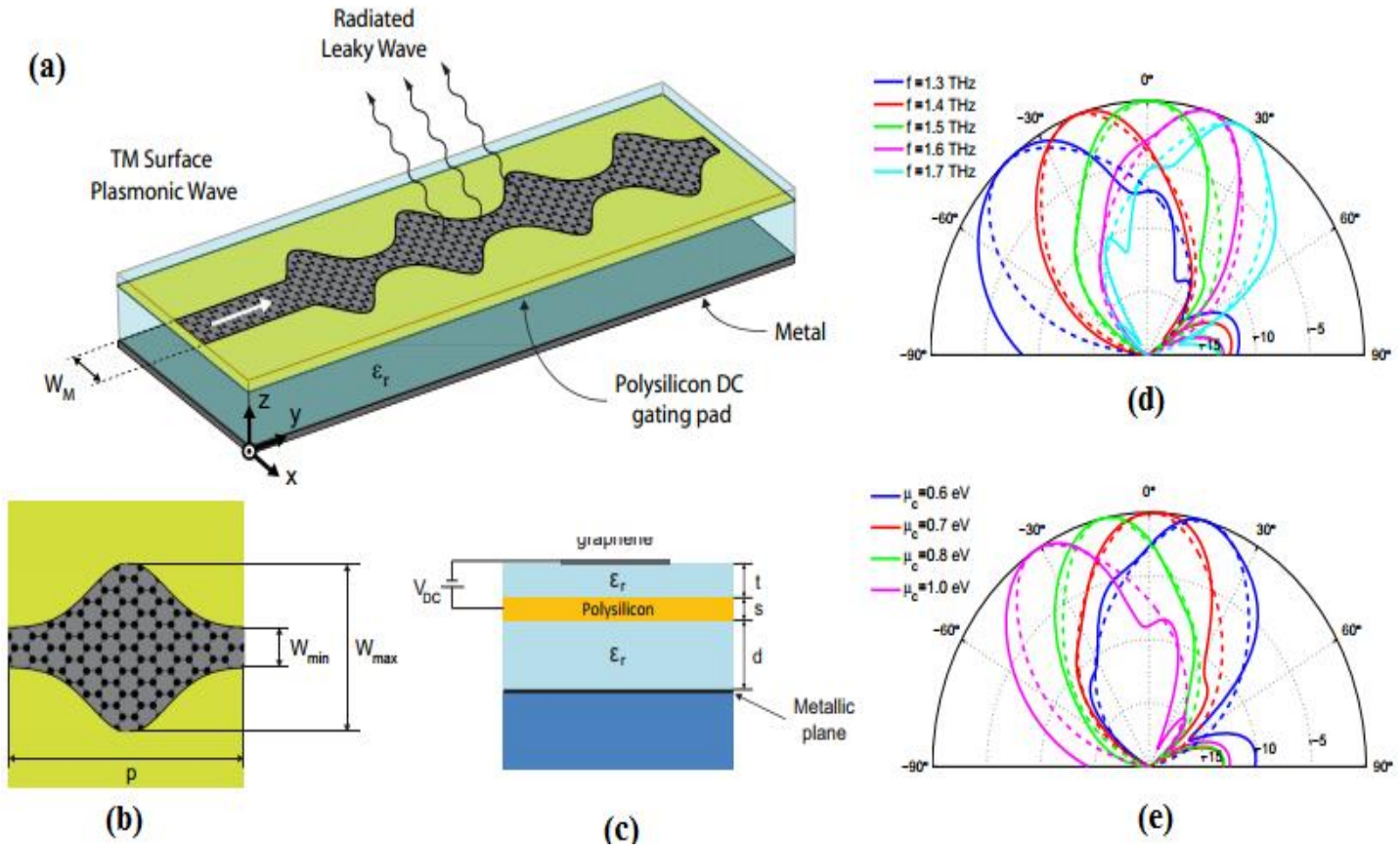

Fig. 23. (a) the leaky wave antenna proposed by graphene strip, (b) top view of unit cell, (c) cross section, (d) radiation patterns for various frequencies for $\mu_c = 0.7\,ev$, (e) radiation patterns for various chemical potentials for $f_0$*=1.5 THz* [94].

In [95], a transmission line model has been derived for SPP on graphene strip, similar to [52]. This model gives a physical insight to graphene strip and consists of resistance, electrostatic capacitance, kinetic and faraday inductance [95]. A similar article such [92] is published which reports the modal properties of individual and paired GNRs used in VLSI optoelectronics [96]. Even and odd modes of coupled GNRs has been presented in in [97, 98], similar to [92, 96]. The authors suggest an equivalent circuit for coupled GNRs based on Method of Moments (MoM) where the graphene is considered as simple local conductivity. E-field and H-field patterns of coupled GNRs at *f=1 THz* are displayed in fig. 24. Haowen Hou et al. focus on studying the edge modes of GNR waveguides [99]. Edge modes are lowest loss SPPs in GNRs where the energy is concentrated on the edge. This research reports edge plasmon properties at first and then investigate the cut-off behavior of edge plasmons in GNRs in detail [99]. A mode analysis of plasmonic waves for graphene nano ribbon waveguides is done in [100], which this article is similar to the paper published in 2011 [92]. In this structure, a graphene core layer (3 nm) with higher chemical potential (0.9 ev) is located in center and cladding graphene layer has lower chemical potential (0.3 ev), as seen in fig. 25. The mentioned structure can be utilized in future plasmonic devices.

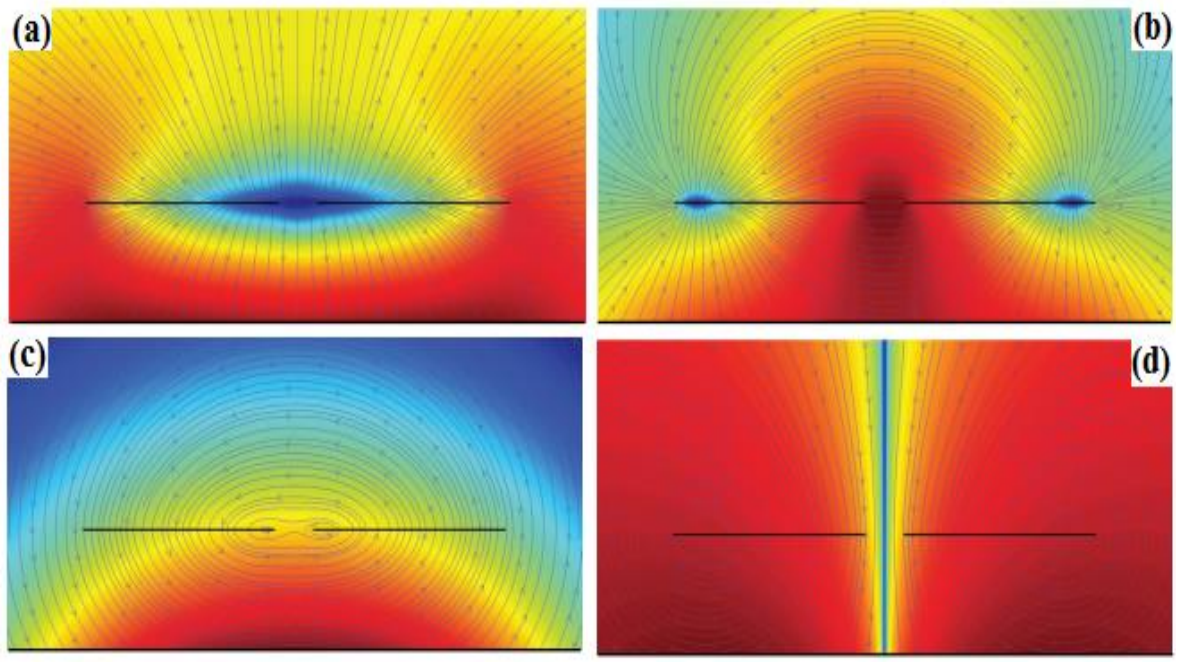

Fig. 24. E-field patterns of: (a) even, (b) odd mode, H-field patterns of: (c) even, (d) odd mode, at *f=1 THz* [97]

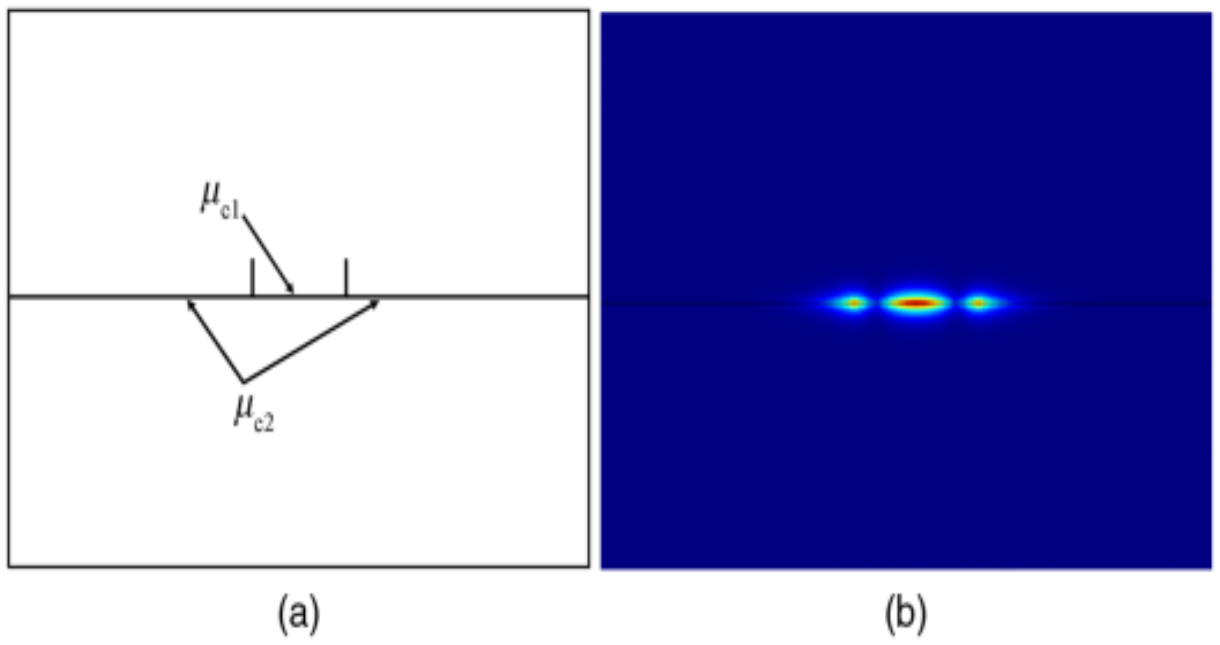

Fig. 25. (a) the proposed GNR, (b) $|E|$ field distribution of the embedded waveguide at *f=190 THz* [100]

### *2.1.5 Rectangular waveguides*

In this section, rectangular metallic waveguide with graphene integrated will be investigated. Gennady Shkerdin and his coworkers have been presented and considered two rectangular metallic waveguides with integrated graphene sheet at *0.3 THz* [101, 102]. Fig. 26 illustrates the proposed waveguide where the graphene layer separates two dielectric layers from each other and the walls of the waveguide are metals [101]. The authors begin to write the electromagnetic fields of $TE_{0m}$ modes and use the boundary conditions to derive dispersion equation for the waveguide. Due to graphene layer, the modal properties of the waveguide is tunable with Fermi level of graphene and the Long range modes can convert to Short-range modes and vice versa by changing the graphene conductivity [101]. Phase modulation study of rectangular metallic waveguide with graphene integrated has been done in second article of Gennady Shkerdin and his coworkers for the structure sketched in fig. 27 [102]. In this waveguide, the graphene is placed on the $SiO_2$ layer ( $w = 0.3\,\mu m$ ). The authors reported linear phase shifts of range 287-340 GHz by varying the chemical potential in the range 0-0.5 ev where the maximum phase shift is achieved 52 degrees at 287 GHz [102].

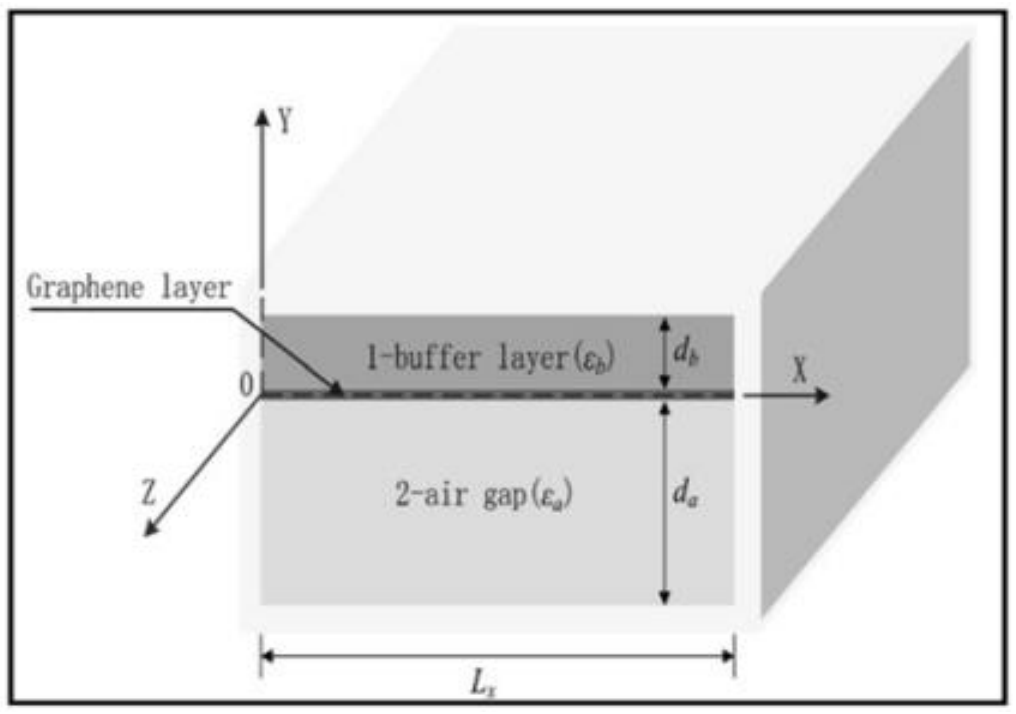

Fig. 26. The rectangular metallic waveguide with integrated graphene layer [101]

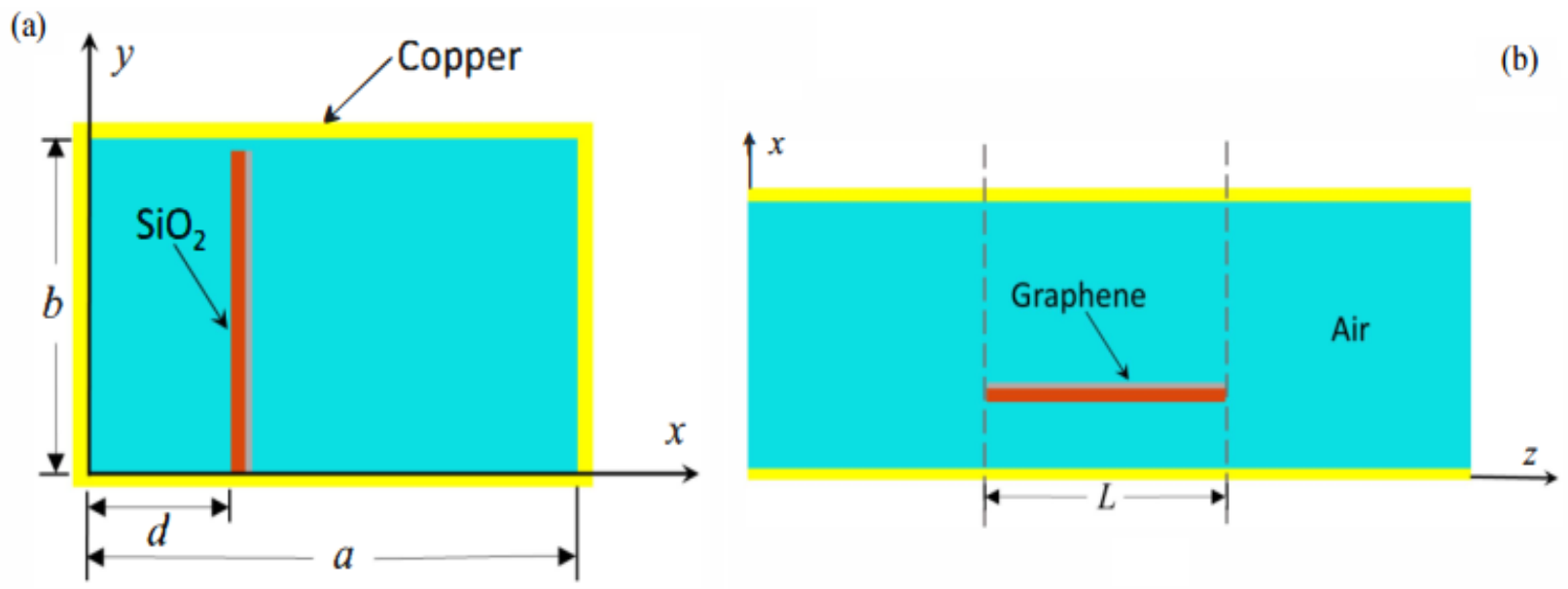

Fig. 27. (a) Cross section, (b) Top view of the rectangular metallic waveguide with integrated graphene. The geometrical parameters are *a=2b=0.864 mm*, *L=0.51 mm*, *d= 0.26 mm* and small gap to upper wall is $g = 0.3\,\mu m$ [102]

## 2.2 Rectangular Graphene Waveguides with Anisotropic Graphene sheet

In this section, rectangular graphene waveguides have been discussed in detail, where graphene sheet is electrically and magnetically biased in all of these structures. As seen in Fig.1, these structures can be categorized to planar, strip and rectangular waveguides. Fig. 28 represents number of publications for each kind of various waveguides when the graphene sheet is magnetically biased.

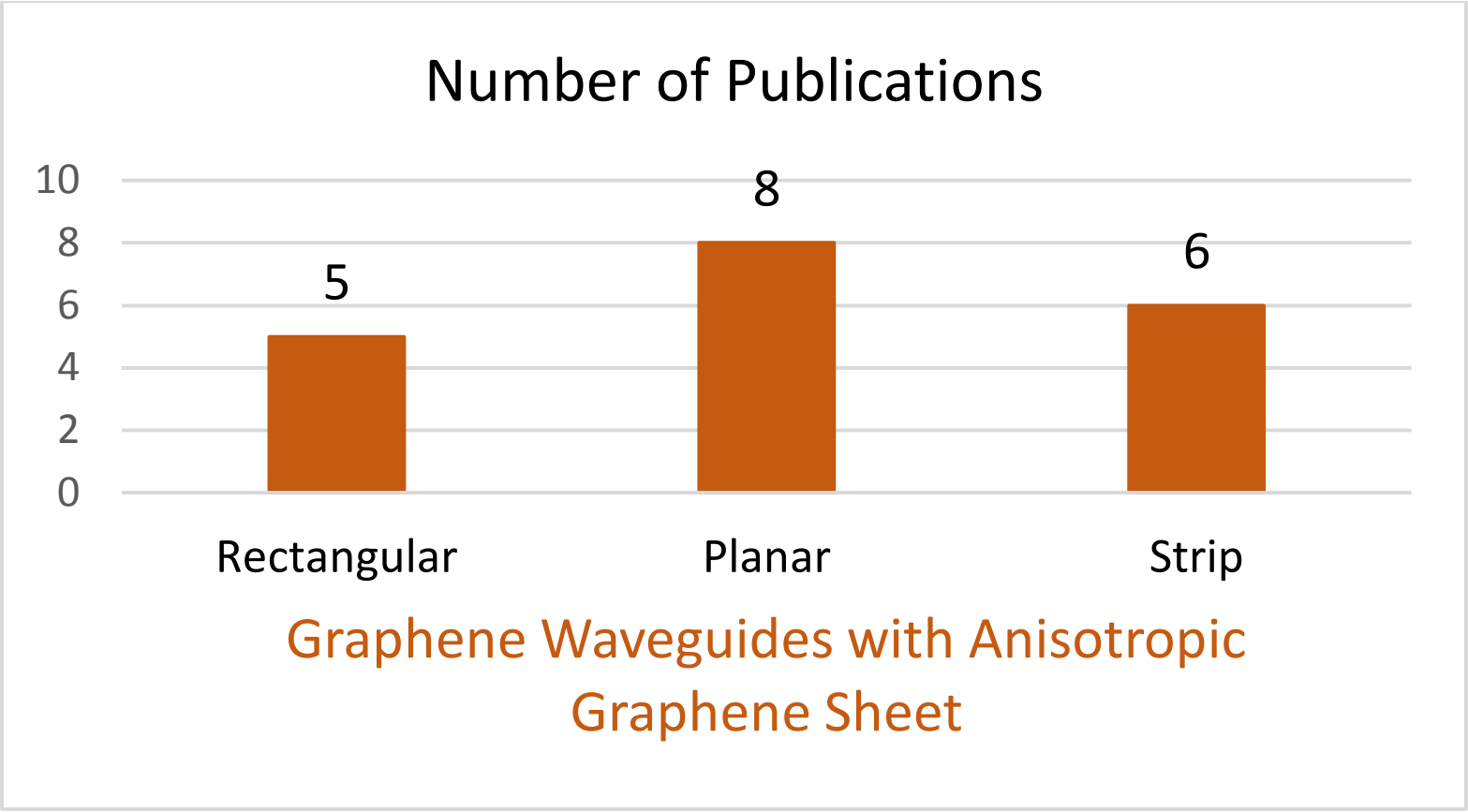

Fig. 28. Number of publications on rectangular graphene waveguides with electrically and magnetically biased graphene from 2007 to 2017

All of these structures use following equation for conductivity of graphene [40]:

$$\overline{\overline{\sigma}}\left(\omega,\mu_c,\Gamma,\mathrm{T},\vec{\mathrm{B}}_0\right)=\begin{pmatrix}\sigma_d & -\sigma_o\\ \sigma_o & \sigma_d\end{pmatrix} \tag{5}$$

Where $\sigma_d,\sigma_o$ are [40]:

$$\begin{aligned}\sigma_d\left(\mu_c,B_0\right)=&\frac{e^2v_f^2\left|eB_0\right|\left(\omega-j2\Gamma\right)\hbar}{-j\pi}\times\\ &\sum_{n=0}^{\infty}\left[\frac{f_d\left(M_n\right)-f_d\left(M_{n+1}\right)+f_d\left(-M_{n+1}\right)-f_d\left(-M_n\right)}{\left(M_{n+1}-M_n\right)^2-\left(\omega-j2\Gamma\right)^2\hbar^2}\times\left(1-\frac{\Delta^2}{M_nM_{n+1}}\right)\times\frac{1}{M_{n+1}-M_n}\right.\\ &\left.+\left[\frac{f_d\left(-M_n\right)-f_d\left(M_{n+1}\right)+f_d\left(-M_{n+1}\right)-f_d\left(M_n\right)}{\left(M_{n+1}+M_n\right)^2-\left(\omega-j2\Gamma\right)^2\hbar^2}\times\left(1+\frac{\Delta^2}{M_nM_{n+1}}\right)\times\frac{1}{M_{n+1}+M_n}\right]\right]\end{aligned} \tag{6}$$

$$\begin{aligned}\sigma_o\left(\mu_c,B_0\right)=&\frac{e^2v_f^2\left|eB_0\right|}{\pi}\times\\ &\sum_{n=0}^{\infty}\left[\frac{f_d\left(M_n\right)-f_d\left(M_{n+1}\right)-f_d\left(-M_{n+1}\right)+f_d\left(-M_n\right)}{\left(M_{n+1}-M_n\right)^2-\left(\omega-j2\Gamma\right)^2\hbar^2}\times\left(1-\frac{\Delta^2}{M_nM_{n+1}}\right)\right.\\ &\left.+\left[\frac{f_d\left(M_n\right)-f_d\left(M_{n+1}\right)-f_d\left(-M_{n+1}\right)+f_d\left(-M_n\right)}{\left(M_{n+1}+M_n\right)^2-\left(\omega-j2\Gamma\right)^2\hbar^2}\times\left(1+\frac{\Delta^2}{M_nM_{n+1}}\right)\right]\right]\end{aligned} \tag{7}$$

Where

$$f_d\left(M_n\right)=\frac{1}{1+\exp\left(\frac{\varepsilon-\mu_c}{K_BT}\right)} \tag{8}$$

and

$$M_n=\sqrt{\Delta^2+2nv_f^2\left|eB_0\right|\hbar} \tag{9}$$

In above equations, $e$ is the charge of an electron, $\hbar$ is the reduced Planck's constant, $K_B$ is Boltzmann's constant, $T$ is temperature, $\omega$ is radian frequency, $\mu_c$ is chemical potential, $v_f \approx 10^6\ m/s$ is the electron velocity in graphene, $B_0$ is applied magnetic field and $\Delta$ is an excitonic energy gap. Relations (6) and (7) become simple if $\mu_c \gg \hbar\omega_c$ and $\mu_c \gg K_B T$ [103]:

$$\sigma_d\left(\omega,\mu_c,\tau,\mathrm{T},\mathrm{B}_0\right)=\sigma_0\frac{1+j\omega\tau}{\left(\omega_c\tau\right)^2+\left(1+j\omega\tau\right)^2} \tag{10}$$

$$\sigma_o\left(\omega,\mu_c,\tau,\mathrm{T},\mathrm{B}_0\right)=\sigma_0\frac{\omega_c\tau}{\left(\omega_c\tau\right)^2+\left(1+j\omega\tau\right)^2} \tag{11}$$

Where

$$\omega_c=\frac{eB_0v_f^2}{|\mu_c|} \tag{12}$$

is cyclotron frequency and the static conductivity for $B_0 = 0$ is [103]:

$$\sigma_0=\frac{e^2\mu_c\tau}{\pi\hbar^2} \tag{13}$$

Where $\tau$ is scattering time which is defined as [103]:

$$\tau=\frac{\pi\hbar^2 n_s\mu}{e\mu_c} \tag{14}$$

where $\mu$ is the DC mobility of graphene and $n_s$ is the surface carrier density.

### 2.2.1 Rectangular waveguides

Jing chen and his coworkers were designed and published first rectangular waveguide with anisotropic graphene sheet in 2013 [104, 105]. In both structure proposed by them, graphene layer are located perpendicular to their axis and magnetically biased with $B_0$. The authors are derived transmission and reflection coefficients of both structures by using the coupling matrix (M-matrix) and normalized transverse electric field Eigen functions [104, 105]. A complete discussion for influence of biased magnetic field and chemical potentials on S-parameters results have been done in these articles [104, 105].

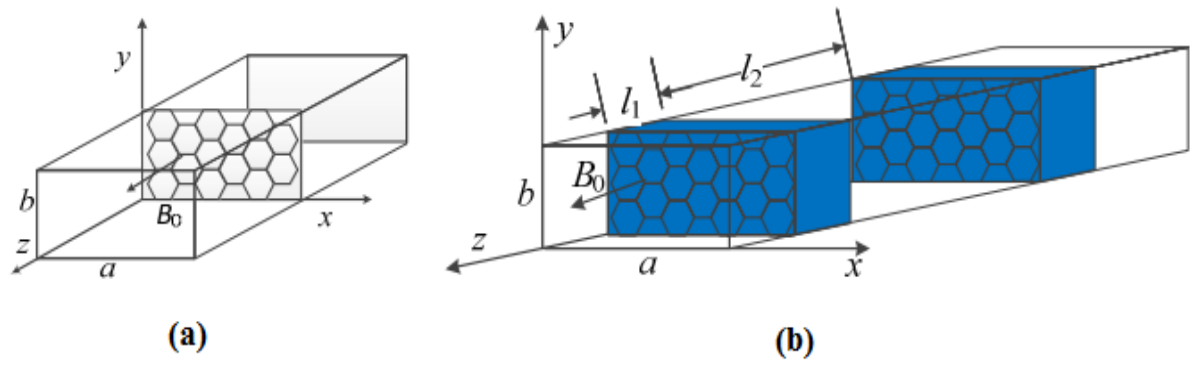

Fig. 29. Rectangular graphene waveguides with: (a) a graphene sheet located perpendicular to its axis[104], (b) multiple graphene sheets with quartz substrates where are placed in distance of $l_2$ from each other [105]

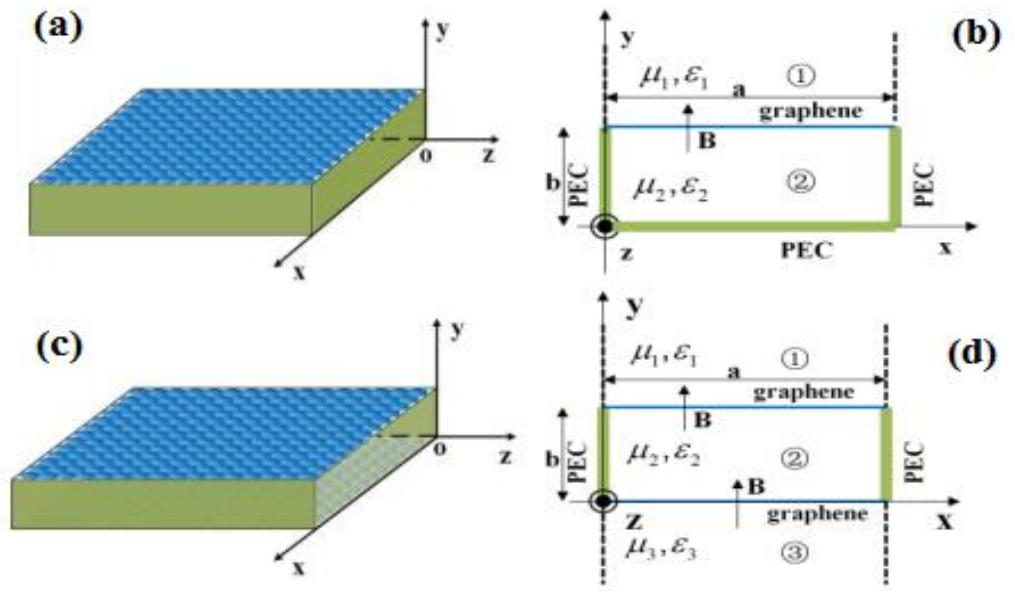

Fig. 30. (a)-(b) 3D view and cross section of rectangular waveguide based on graphene-PEC, (c)-(d) 3D view and cross section of rectangular waveguide based on graphene-graphene [106]

Another group research in shanghai Jiao Tong university are studied graphene based waveguides with anisotropic graphene sheet [106-108]. In first structure, the graphene layer is placed on top of the rectangular waveguide and other sides are PEC whereas two graphene layers are placed in upper and lower plates of waveguides in second structure, as illustrated in fig. 30 [106]. By using Maxwell's equation and conventional relations for hertz electric and magnetic potentials, complicated dispersion relations are achieved and depicted for both structures. Due to anisotropic graphene sheet, the hybrid modes are existed at frequencies beyond their cut-off frequency in both structures and can be tuned by external magnetic field [106-108].

### *2.2.2 Planar waveguides*

J. Perruisseau-Carrier and his coworkers have investigated the hybrid SPPs propagating in the anisotropic graphene sheet sandwiched between two dielectrics, as illustrated in fig. 31 [109, 110]. In this structure, the graphene sheet is magnetically biased in perpendicular direction (*z*-axis) and the whole waveguide is illuminated by a plane wave with arbitrary direction [109]. The authors have proposed the transmission line model for this waveguide, as seen in fig. 31 (b).Properties of hybrid TE-TM plasmons on magnetically bias graphene are studied and results showed that combination of electrostatic and magneto static biasing fields can control and enhance the field localization (in comparison with SPPs on metals) and its location frequency [109, 110]. This field localization can be used in designing miniaturized and tunable plasmonic devices such as sensors and non-linear structures [110].
First parallel plate waveguide with anisotropic graphene plate is introduced by S.Ali Malek abadi and his co-workers in [111] and developed for studying more precisely in 2013 [112].They considered two parallel plate waveguides based on isotropic graphene layer, as exhibited in fig. 32 [112]. In first structure, PEC and graphene layer has been placed at $y=0$ and $y=d$, respectively, whereas in second structure, another graphene layer has been located at $y=0$ and Graphene layers in both structures are electrically and magnetically biased [112]. The hybrid electromagnetic fields have been written in various regions of both structures and then the complicated dispersion relations for both waveguides are obtained by applying boundary conditions [112]. Real and imaginary parts of propagating constants of first structure are depicted in fig. 33 for various chemical potentials and magnetic fields. One can see that the propagation of hybrid SPPs can be controlled by the magnetic bias and chemical potential and changing the magnetic bias has great influence on the propagation characteristics of propagating waves[112].

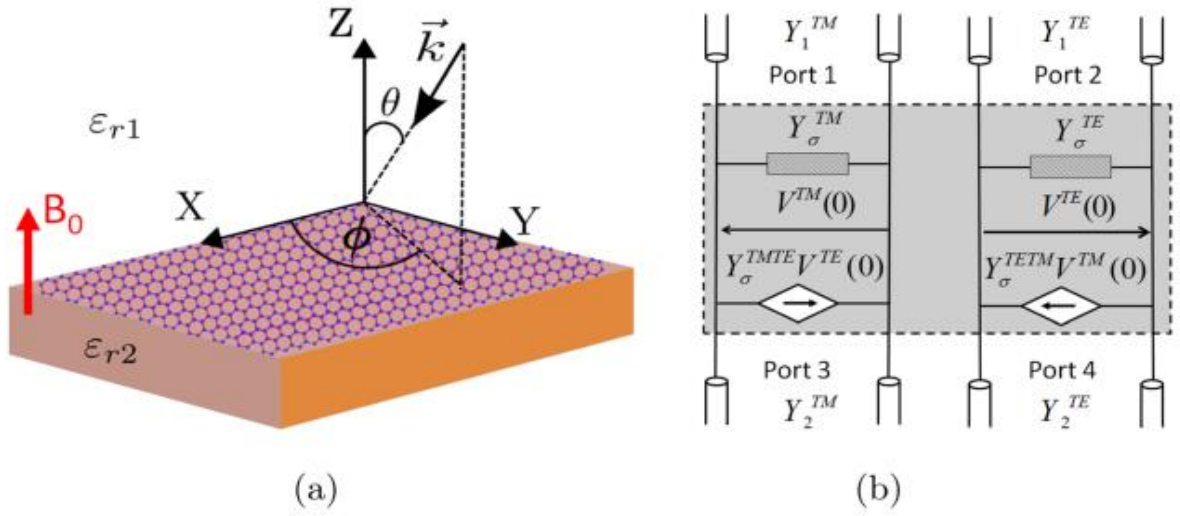

Fig. 31. (a) The isotropic graphene sheet sandwiched between two dielectrics where illuminated by a plane wave, (b) Equivalent transmission line model [109]

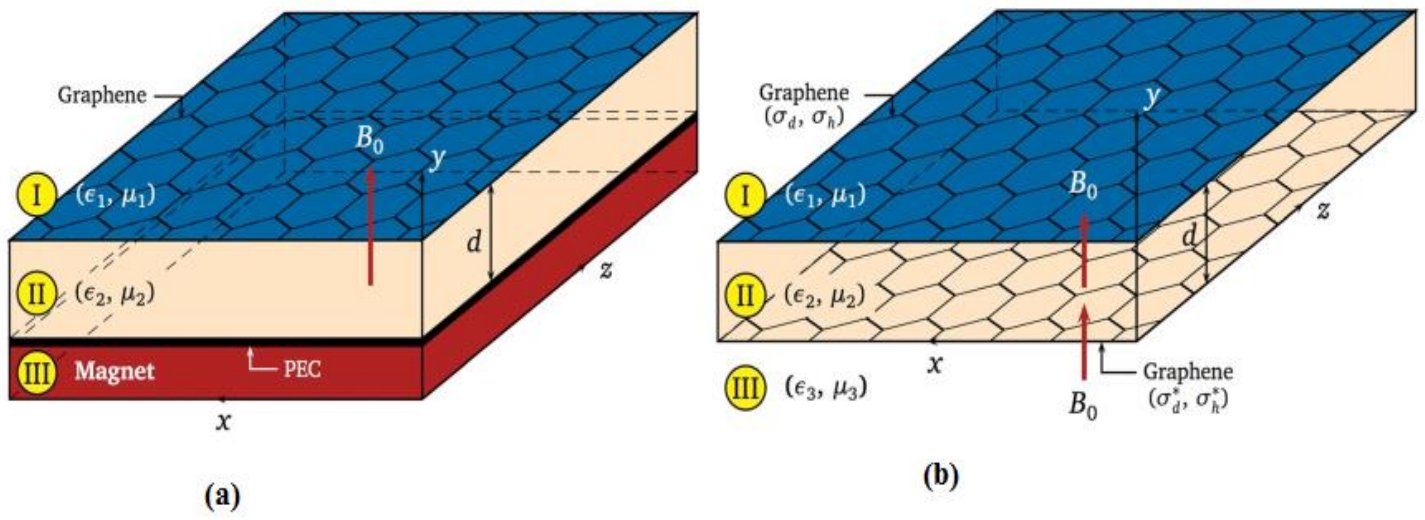

Fig. 32. The parallel plate waveguides: (a) PEC layer and anisotropic graphene sheet are located at $y=0$ and $y=d$, respectively, (b) Two different anisotropic graphene sheets are located at $y=0,d$ [112]

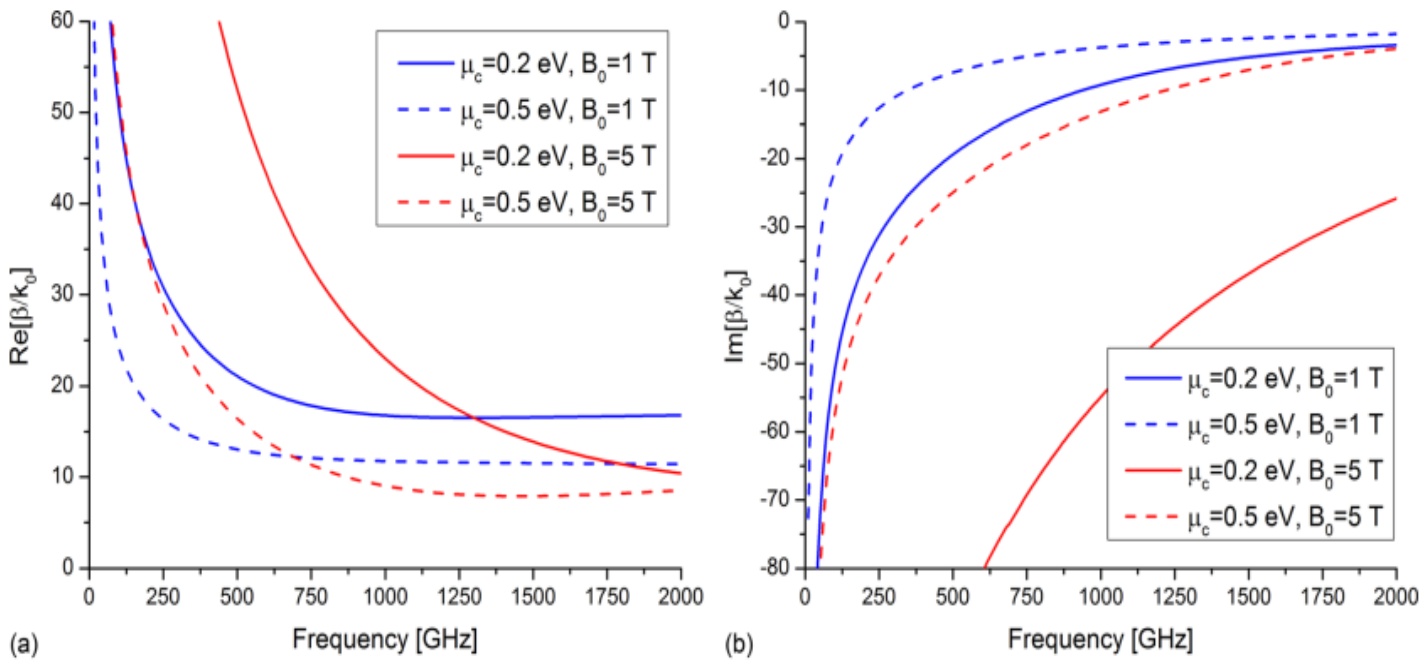

Fig. 33. Real and imaginary parts of propagation constants of hybrid SPPs for first structure of fig. 32 with *d=100 nm*, for various magnetic bias fields and chemical potentials [112]

Another researchers tried to present new aspects of plasmonic waveguides based on anisotropic graphene sheet illustrated in figures 32, 33. For instance, a similar work to [109] for studying the hybrid surface SPPs supported by an isotropic graphene layers sandwiched between two dielectrics (see fig.32) has been done by I.V.Irash et al. [113]. In this article, authors have achieved the complicated dispersion relations by applying boundary conditions instead of the transmission line model achieved in [109]. In [114], a new investigation for hybrid SPPs of structure fig. 32(b) has been reported by Bin Hu and his coworkers. This article gives similar relations for the mentioned structure as obtained in [112]. A coplanar waveguide with an isotropic graphene layer is proposed in [115]. The authors used the FDTD method to simulate the coplanar waveguide to discuss about the odd and even modes propagating in the structure [115]. In [116], multi-layer structure composed by N graphene layers which are located in the borders of various dielectrics is analytically studied by applying transfer matrix method for optical excitation illuminated by an arbitrary TM polarized wave. The reflection spectrum shows three dips which are due to the excitation of hybrid SPPs [116].

### *2.2.3 Strip waveguides*

First study on edge modes of graphene strips where graphene is magnetically biased is reported by Dimitrios L.Sounas and Christophe Caloz in [103]. In this work, a finite graphene strip with width of *w* is magnetically biased $B_0$ and the dispersion diagrams are represented in fig. 34. It is obvious that two types of SPP modes called edge and bulk modes are existed and slow-wave factor of edge modes are greater that slow-wave factor of infinite layer mode while bulk modes have smaller slow-wave factors [103]. The authors place a PEC plate in one of the edges which can short the SPPs in that edge and this feature can be utilized in designing non-reciprocal devices such as isolators [103].

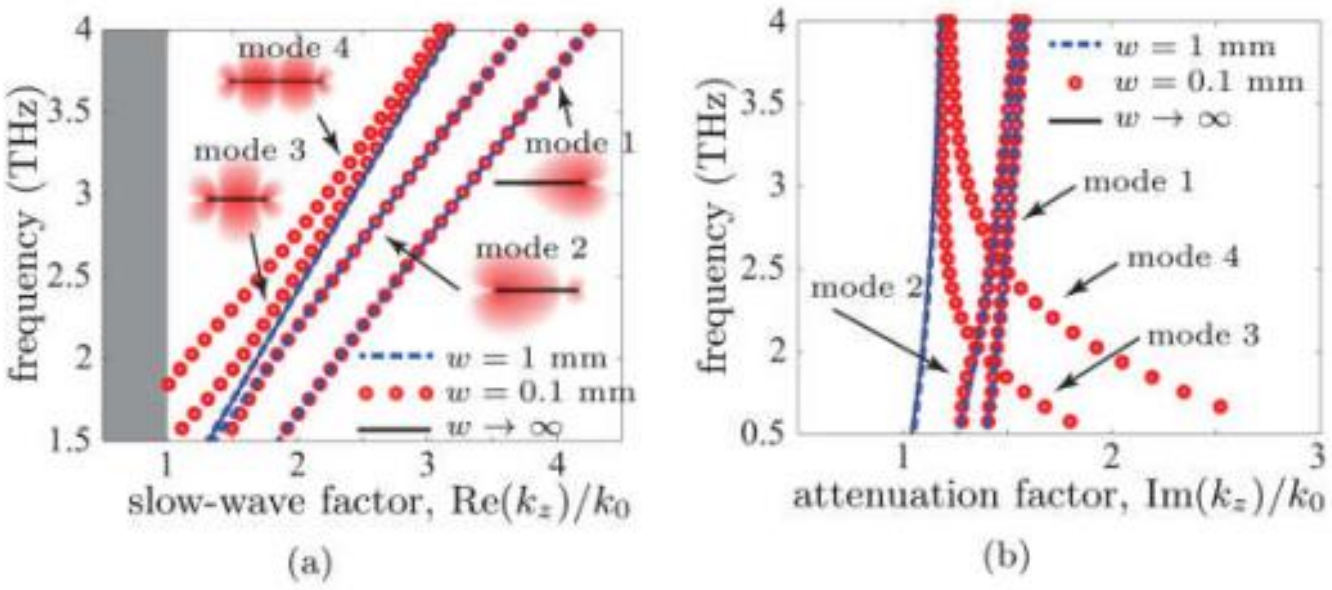

Fig. 34. (a) Slow wave factor, (b) Attenuation factor of graphene strip with different width for $B_0$= *1T*, the electric field distributions for three modes are displayed at *f=2 THz* [103]

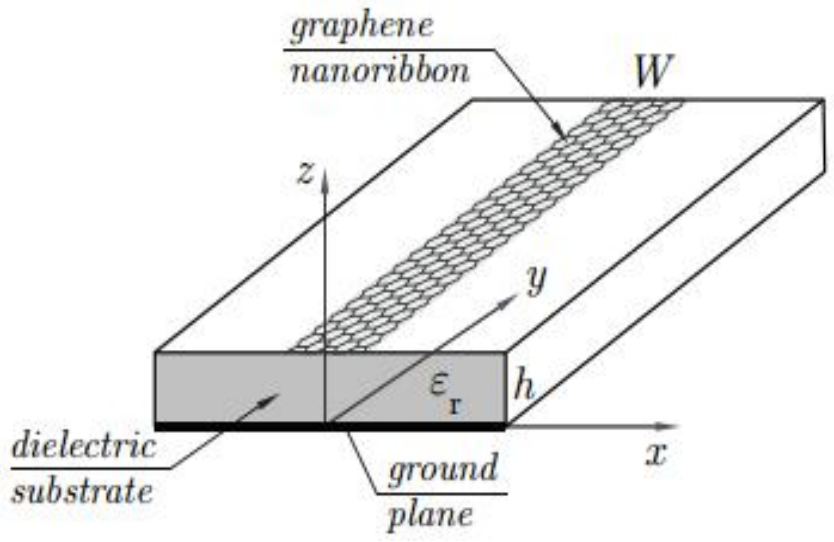

Fig. 35. The GNR strip placed on the grounded dielectric [117]

Giampiero Lovat and his coworkers have analyzed an anisotropic GNR with width *w* deposited on the grounded dielectric for the first time [117-119]. The analysis is done in spectral domain starting by electric field integral equation and using the MoM procedure [119]. The full discussion on numerical results of SPP characteristics such as normalized propagation constants, modal currents and field distributions are given in the article which confirm that spatial dispersion must be account through the model to obtain accurate results [119].
In [120], a circuit model for magnetized graphene sheet by using Kubo formula (see equations (6),(7)) is introduced. The more details has been explained in the article [120]. Similar to [117-119], a microstrip transmission line with the anisotropic graphene layer has been investigated by spectral domain method and method of lines [121]. The authors concluded that the characteristics impedance of the structure can be tuned by chemical potential and magnetic bias of graphene which distinguishes it from conventional micro strips in Microwave regime [121].

## 3. CLYNDERICAL GRAPHENE WAVEGUIDES

In this section, various cylindrical waveguides have been introduced and discussed in detail. These graphene waveguides can be divided to two main categories: 1- Graphene waveguides with electrically biased graphene.

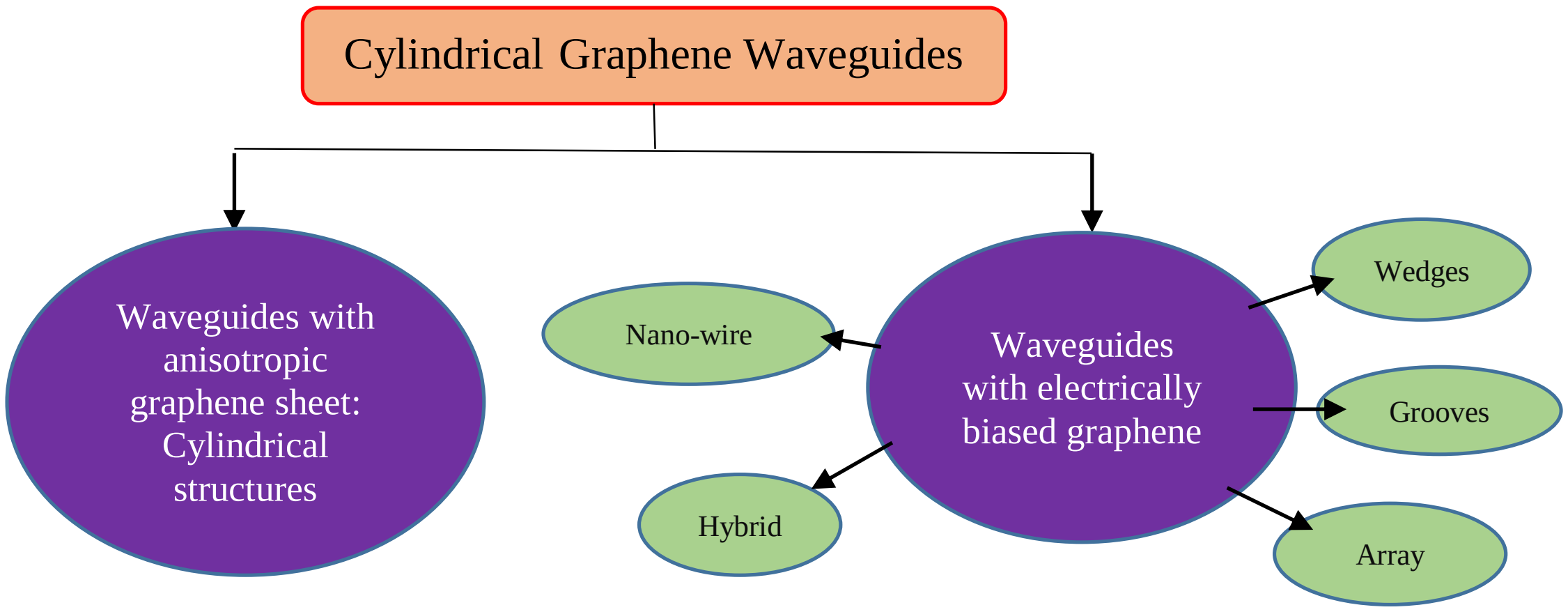

Fig. 36. Cylindrical graphene waveguides

2-Graphene waveguides with anisotropic graphene sheet (magnetically biased graphene). In Fig. 36, various cylindrical graphene waveguides have been illustrated. This section gives a historical review to cylindrical graphene waveguides.

### 3.1 Cylindrical Graphene Waveguides with Electrically Biased Graphene

There are numerous articles that have considered the various cylindrical graphene waveguide structures. In this section, cylindrical graphene waveguides have been discussed in detail, where graphene sheet is electrically biased in all of these structures. As seen in Fig.36, these waveguides can be categorized to various structures such nano-wire, hybrid, wedges and grooves, etc. All of these structures use Kubo equation for graphene sheet which is expressed in relation (1) in previous section. Fig. 37 represents number of publications for each kind of various waveguides. It is obvious that nano-wire graphene waveguides have maximum publications among other types of waveguides.

#### *3.1.1 Nano-wire waveguides*

Studying SPPs in graphene-coated nano-wires is one of the interesting subjects which has been done by researchers in recent years. In this section, a brief historical review of these kind of waveguides will be presented. Yizhe Yaun et al. focused on considering the photonic states of graphene-coated cylindrical structure as photonic crystal in [122]. In this structure, area I and III are various dielectric layers and graphene layer is located on area II ($\rho = \rho_0$), as exhibited in fig. 38 (a). By starting the wave equations and applying boundary conditions, complicated dispersion relation can be obtained and numerically solved, as demonstrated in fig. 38 (b) [122]. This dispersion diagram indicates that as frequency increases, the wave number increases for different modes.

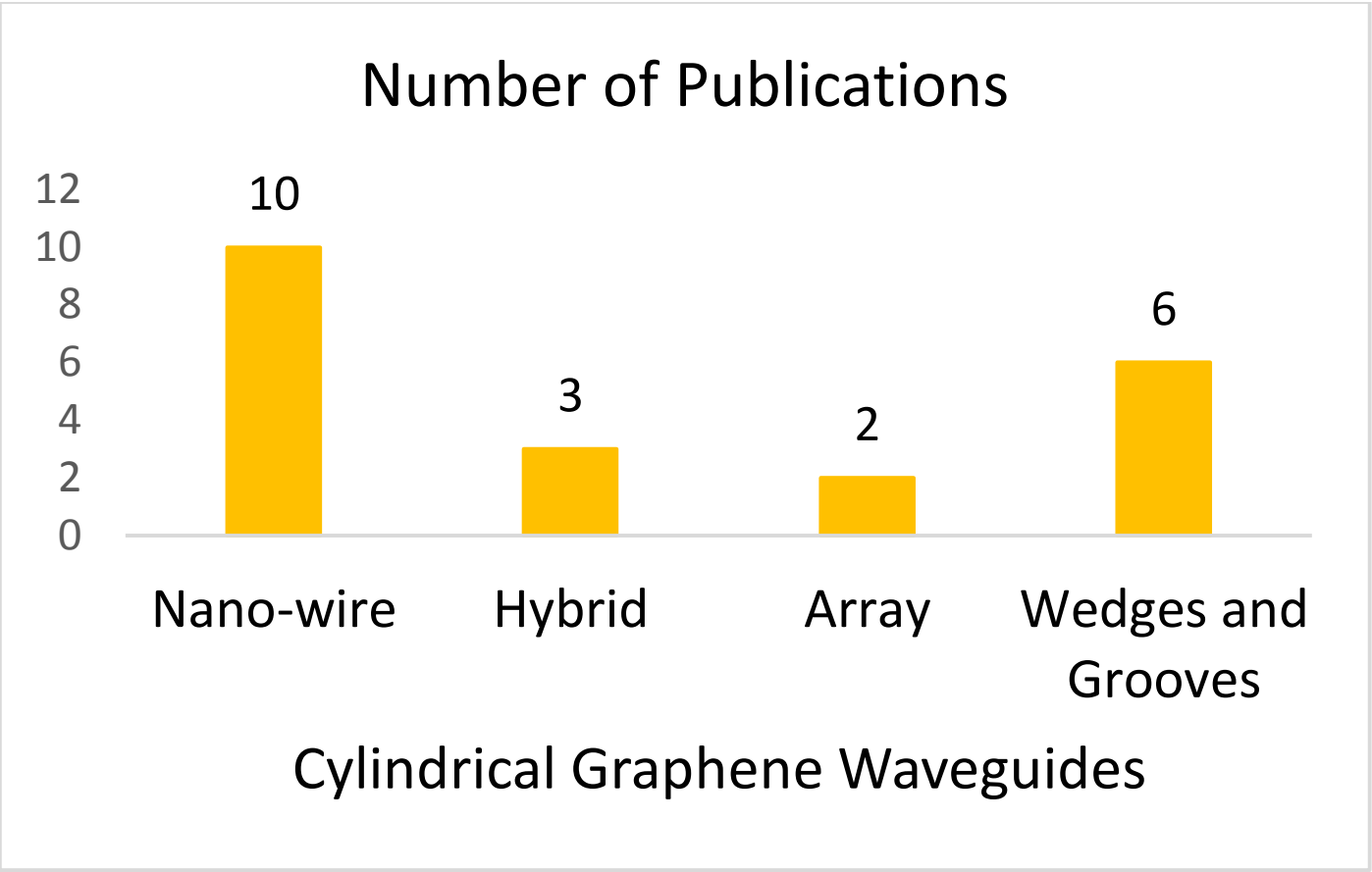

Fig. 37. Number of publications on cylindrical graphene waveguides with electrically biased graphene from 2007 to 2017

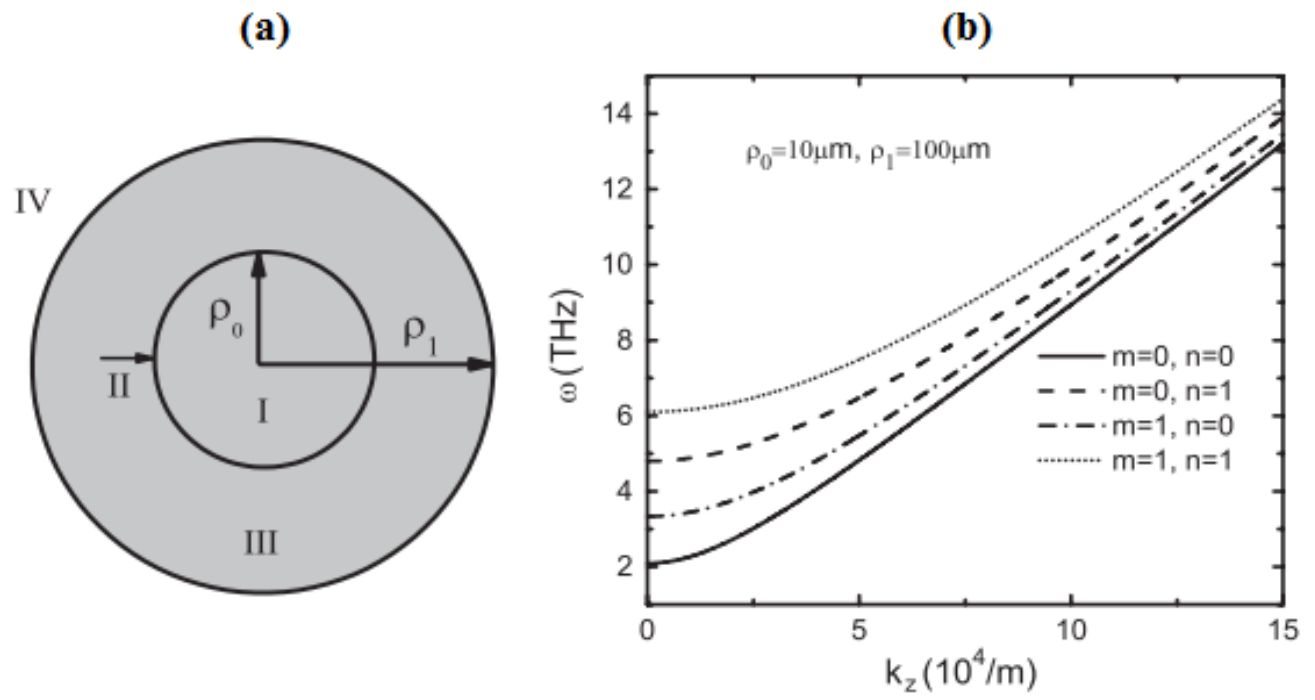

Fig. 38. (a) Graphene based nano-wire used as photonic crystal: the graphene layer (area II) is located between two dielectric cylindrical layers (area I and III), (b) dispersion diagram for $\rho_1 = 10\rho_o = 100\ nm$ [122]

In [123-125], SPPs in graphene cylindrical waveguides are studied more precisely. In [123], a dispersion relation is achieved for cylindrical graphene waveguide where the graphene is placed on dielectric surface with radius *R*. A more detailed study for graphene coated nano-wires has been investigated by Yixiao Gao and his co-workers in [124, 125]. The schematic of graphene coated nano-wire studied in these articles has been displayed in fig. 39 [125]. By using Maxwell's equations in cylindrical coordinates, the longitude components of fields can be written in two regions and then the dispersion equation can be derived by applying boundary condition at $\rho = R$ [125]. The structure is simulated and field distributions, dispersion diagram and propagation length are plotted for first five order modes versus frequency [125]. It is obvious that mode $m=0$ is cut-off free and effective mode indices of other modes decrease as frequency reduces. In the rest of paper, a comparison of SPP modes between metal coated and graphene coated nano-wire has been done [125]. The authors have concluded that graphene nano-wire has higher mode index and better mode confinement which can be used in many applications such as sensing and integrated plasmonic circuits [125].

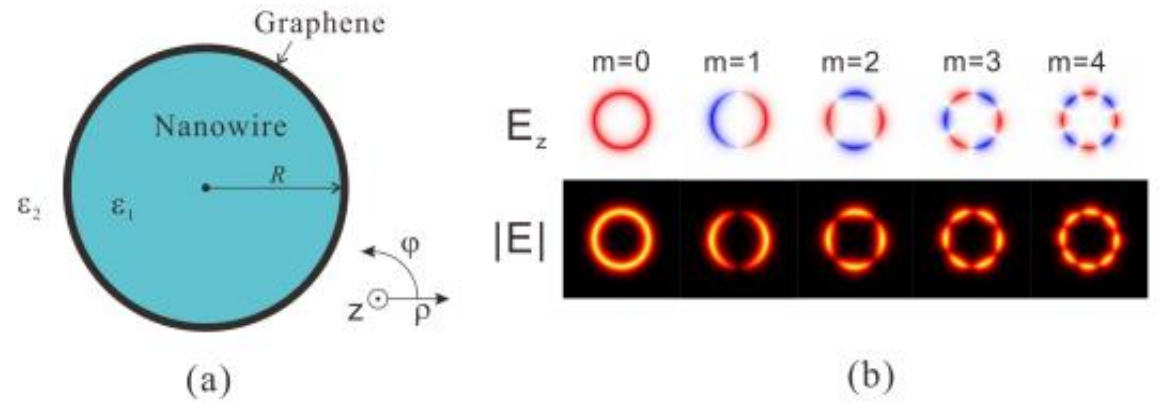

Fig. 39. (a) Graphene nano-wire where graphene is located between two dielectric cylindrical layers, (b) field distribution of first five order modes at *50 THz* [125]

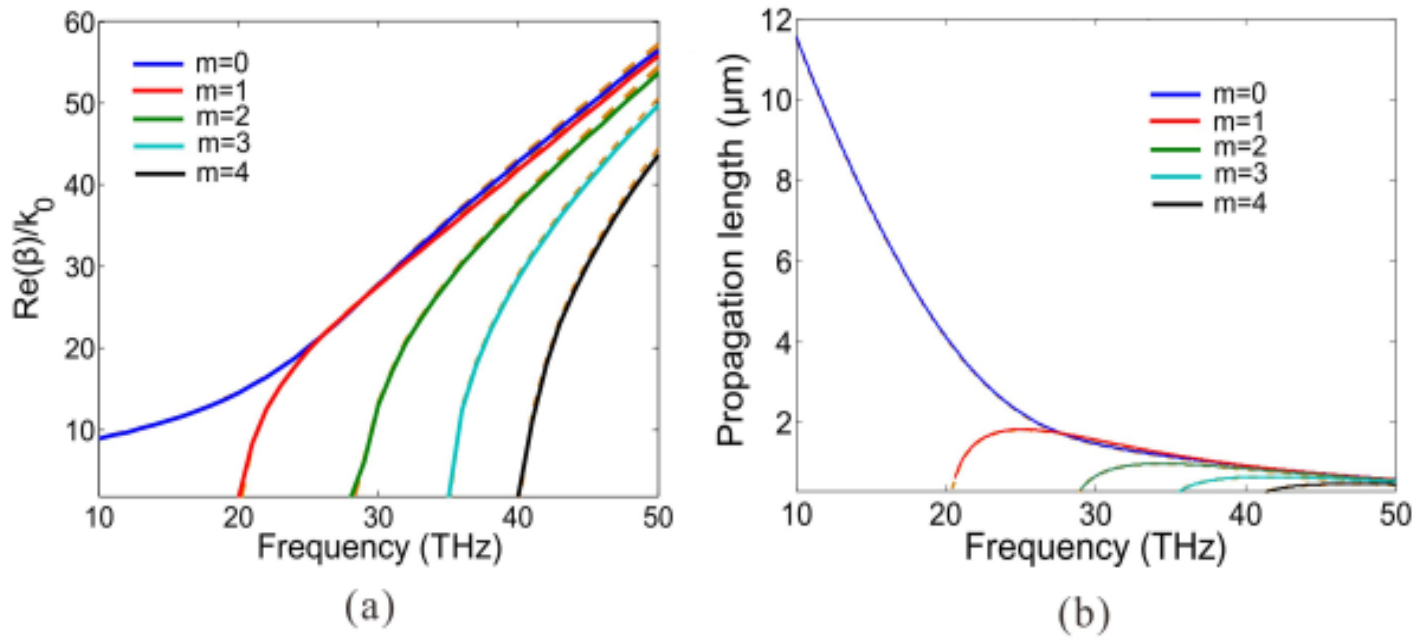

Fig. 40. (a) Dispersion diagram, (b) Propagation length for structure dig. 39 (a) [125]

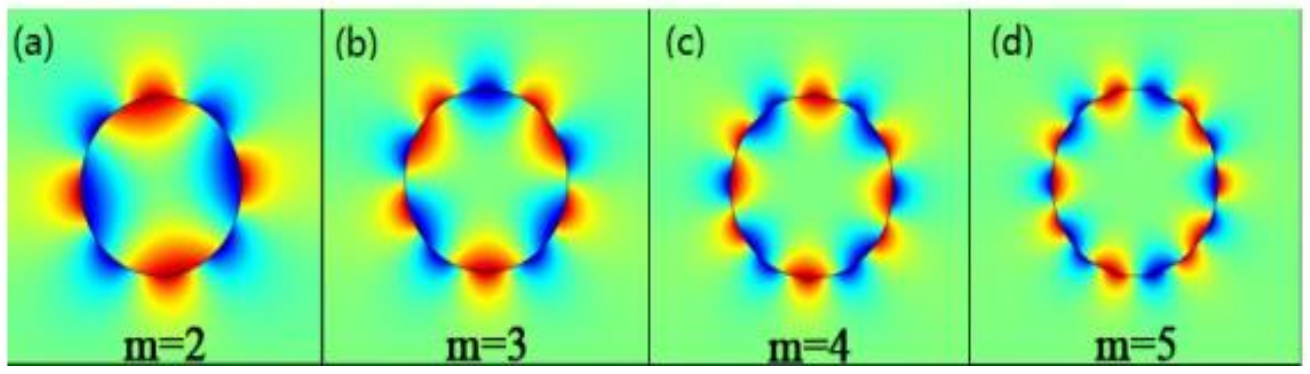

Fig. 41. Electrical field $E_x$ distributions for various whispering-gallery modes for the cavity with radius *5 mm* [126]

Whispering gallery modes are a kind of waves that can propagate in concave surface and has many applications such lasing and sensing. These waves in the resonators can give high quality factor but always has lossy mechanism even in theoretically ideal conditions. The authors in [126] consider SPP whispering gallery modes of graphene monolayer coated on InGaAs cavity. The studied structure in this research is similar to fig. 39 (a) where it is assumed that inner dielectric is InGaAs (n=*3.45*) an outer space is air (*n=1*) [126]. By simulating the structure in the COMSOL software, the electric field $E_x$ distributions with different azimuth number can be demonstrated in fig.41. The authors reported quality factor of 235 for a 5 nm radius cavity with mode area $3.75\times10^{-5}\ \lambda_0^2$ [126].

The study on cylindrical graphene nano-wires has an enhancement in 2016. For instance, Vladimir Shavrov and his coworkers have been published two research articles in this year [127, 128]. Magneto optical faraday effect occurs when the polarization of the light changes in magnetic medium. In the inverse situation, the polarized photons can be affected on magnetization which called Inverse Faraday Effect (IFE) and this phenomena can be used to design novel photonic devices. In [127], a graphene nano-wire waveguide has been analyzed theoretically to consider IFE, which the mentioned structure is similar to structure of fig. 38 (a) [122]. The authors result that basic mode can be induce vertex-like magnetic distribution and modes with higher modes can induce a longitudinal magnetic components that can be rotates along the axis [127]. A novel article about TE SPPs on the cylindrical graphene based waveguides at near-infrared and visible frequencies has been published by Vladimir Shavrov and his coworkers [128], which has similar structure of fig. 39 (a). In this paper, first authors derived a condition for propagating TE mode at near infrared frequencies and then studied characteristics of guiding TE modes. This is the first paper that studies the TE modes of cylindrical graphene waveguides at infrared frequency and suggests that these TE modes can be applied to design and fabricate optical devices including telecommunication frequencies [128].

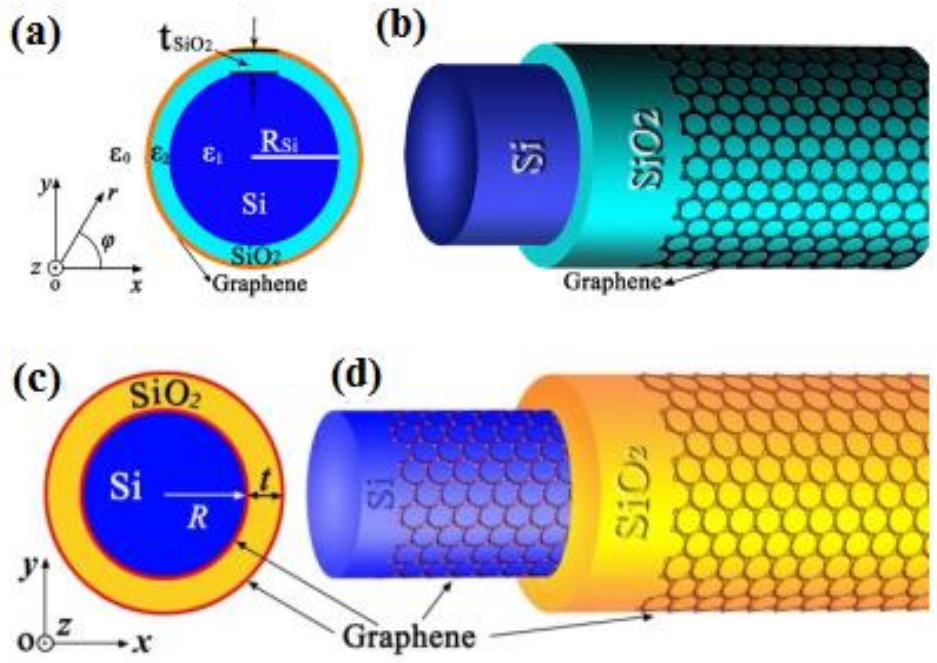

Fig. 42. (a), (b) cross section and 3D view of cylindrical waveguides
with one layer graphene [129], (c),(d) double layer graphene [130]

A graphene based cylindrical waveguides with one layer graphene [129] and double layer graphene [130] are proposed and investigated by Jian-Ping Liu et al, as shown in fig. 42. For both structures, the fields in various regions have been written and dispersion relation can be obtained by using boundary conditions. The modal properties and field distributions of various modes are demonstrated and discussed more precisely. In the last section of [129], the coupling of two graphene based cylindrical waveguides are investigated. The authors use perturbation theory of coupled mode to describe the coupling relations [129]. A similar article for circular cylindrical double layer graphene has been published [131]. In this work, fields are written in various regions and boundary conditions are applied at boundaries, similar to [130].

### *3.1.2 Hybrid waveguides*

In the literature, some articles have been reported hybrid graphene waveguides with cylindrical structures. In [132], authors propose a hybrid graphene waveguide where two cylinders (with radius R) are placed at each side of graphene sheet. The structure is simulated in COMSOL at *10 THz* for geometrical parameters of $R = 2\,\mu m$, $t = 0.7\,nm$, $g = 0.5\,\mu m$ and the permittivity of Si and $SiO_2$ are 12.25 and 2.25, respectively. Electric field distribution shows that the most of energy is concentrated in the gap between two cylinders and has small mode area [132]. The authors investigate the tunability of modal properties as the chemical potential is varied and also they study crosstalk between two adjacent hybrid waveguides [132]. Similar to structure fig. 43, Haoliang Qian and his coworkers designed a nanowire hybrid structure, where the graphene sheet is deposited on the $SiO_2$-Si substrates and silver nano-wire is on the top graphene structure [133]. This structure works as modulator near the Dirac point of graphene [133].
Fig. 44 shows a hybrid plasmonic waveguide where consists of graphene coated nano-wire located on top of two-layered $MgF_2$-Si structure [134]. In this structure, $MgF_2$ has been used as buffer layer and silicon is substrate. The authors have done a parametric study to consider the influence of geometrical parameters on propagation length and normalized mode area [134]. The parametric study reveals that high Figure of Merit (FOM=108.1) and long propagation length ($13.91\,\mu m$) can be obtained by choosing G=H=5 nm and $R < 20\,nm$ [134]. It should be noted that most energy is concentrated in G gap.

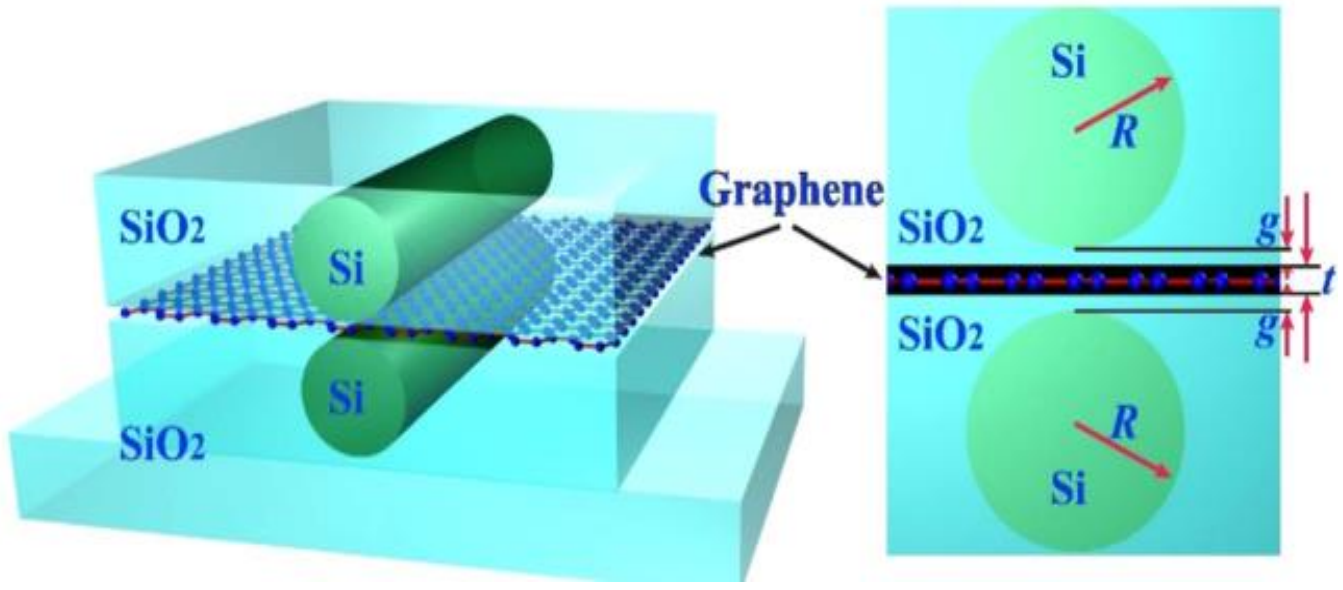

Fig. 43. Hybrid graphene waveguide where two cylinders (with radius R) are placed at each side of graphene sheet [132]

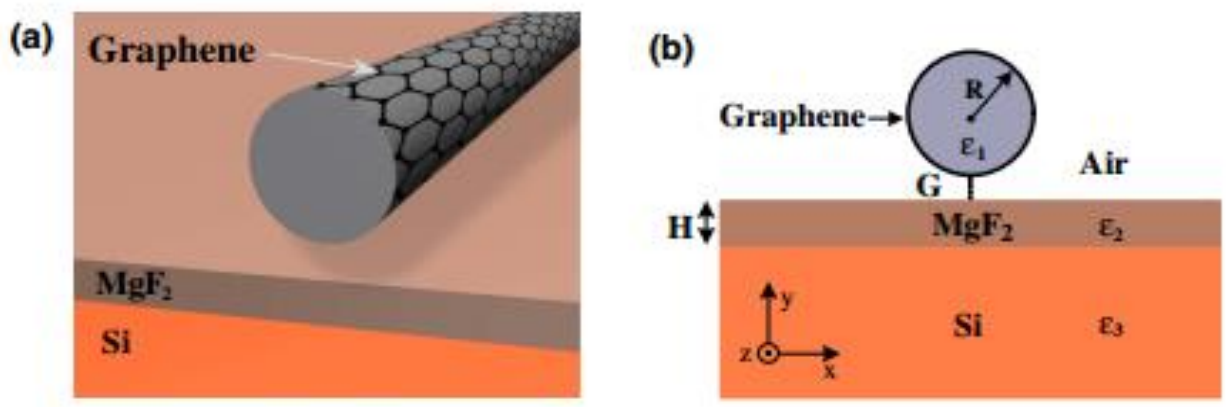

Fig. 44. (a) 3D view and (b) cross section of the hybrid graphene waveguide

### *3.1.3 Nano-wires Array*

Graphene coated nano-wire arrays have been proposed and studied in two articles [135, 136]. In [135], a periodic array of dielectric nano-wires coated by graphene is presented where cylinders with radius a have distance s from each other. The nonlinear conductivity for graphene in mid-infrared spectrum is used and discrete solitons modes are produced whose modal properties can be tuned by graphene conductivity [135]. The authors investigate one-dimensional and two-dimensional of period arrays where the electric field distributions $E_z$ are depicted for one-dimensional case. It should be emphasized that compact and low loss solitons modes with tunability of mode properties by chemical potential can be utilized in non-linear optics to design attractive devices such switches [135].

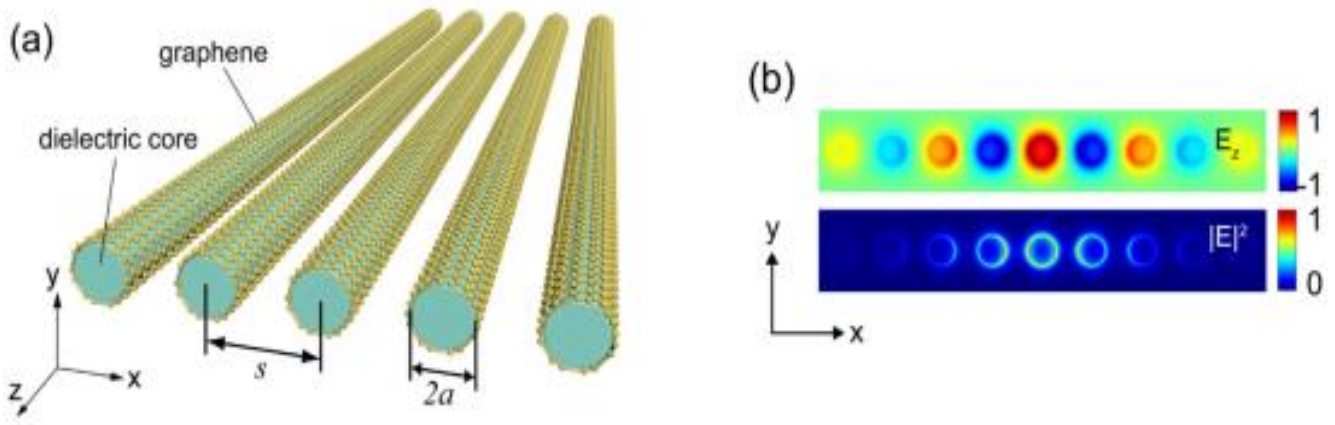

Fig. 45. (a) One-dimensional array of dielectric nano-wires coated by graphene, (b) electric field distributions $E_z$ for graphene-based solitons at $\lambda_0 = 10\,\mu m$ for geometrical parameter $s = 4a = 400\,nm$ and $E_F = 0.5\,eV$, $I_{\max} = 7.2\times10^{15}\ V^2/m^2$ [135]

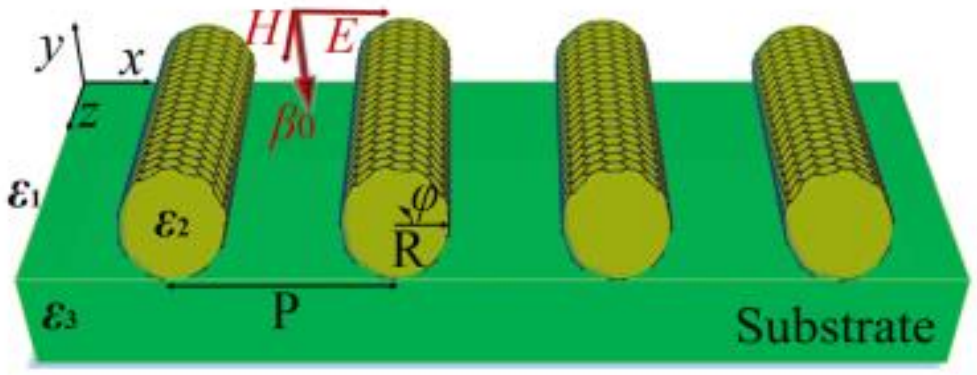

Fig. 46. Graphene-coated nano-wire arrays illuminated with normally incident plane wave [136]

Sheng-Xuan Xia et al. focus on excitation of SPPs in graphene-coated nano-wire arrays [136]. Fig. 46 illustrates the nano-wire arrays where graphene is located on the surface of Indium- Tin- Oxide (ITO) cylinder with permittivity of 2.3 in the dielectric environment of $\varepsilon_1$ and whole array are deposited on the dielectric substrate with permittivity of $\varepsilon_3$ [136]. The authors excite the structure by a normally incident plane wave with *x*-polarized electric field, as displayed in fig. 46. The authors study the proposed array by a parametric study to obtain good physical insight for influence of various parameters on transmission spectrum [136].

### *3.1.4 Wedges and grooves*

Wedges and grooves are interesting waveguides which are studied and used in microwave regime. By starting the great research on graphene since 2007, graphene based wedges and grooves have been considered in some research articles. First study on graphene- based wedge and groove waveguides was done by Penghong Liu, et al. in 2013 [137]. In this paper, dispersion diagrams and field distributions are depicted which SPPs in these waveguides can be categorized into PEC and PMC symmetric modes. Fig. 47 represents the geometry of wedge waveguide and field distributions of $E_y$ for A-D modes [137]. It is obvious that lower order modes A,D presents PEC symmetry while plasmon modes B and C possess the PMC symmetry [137]. The similar study for grooves has been done in the paper [137].

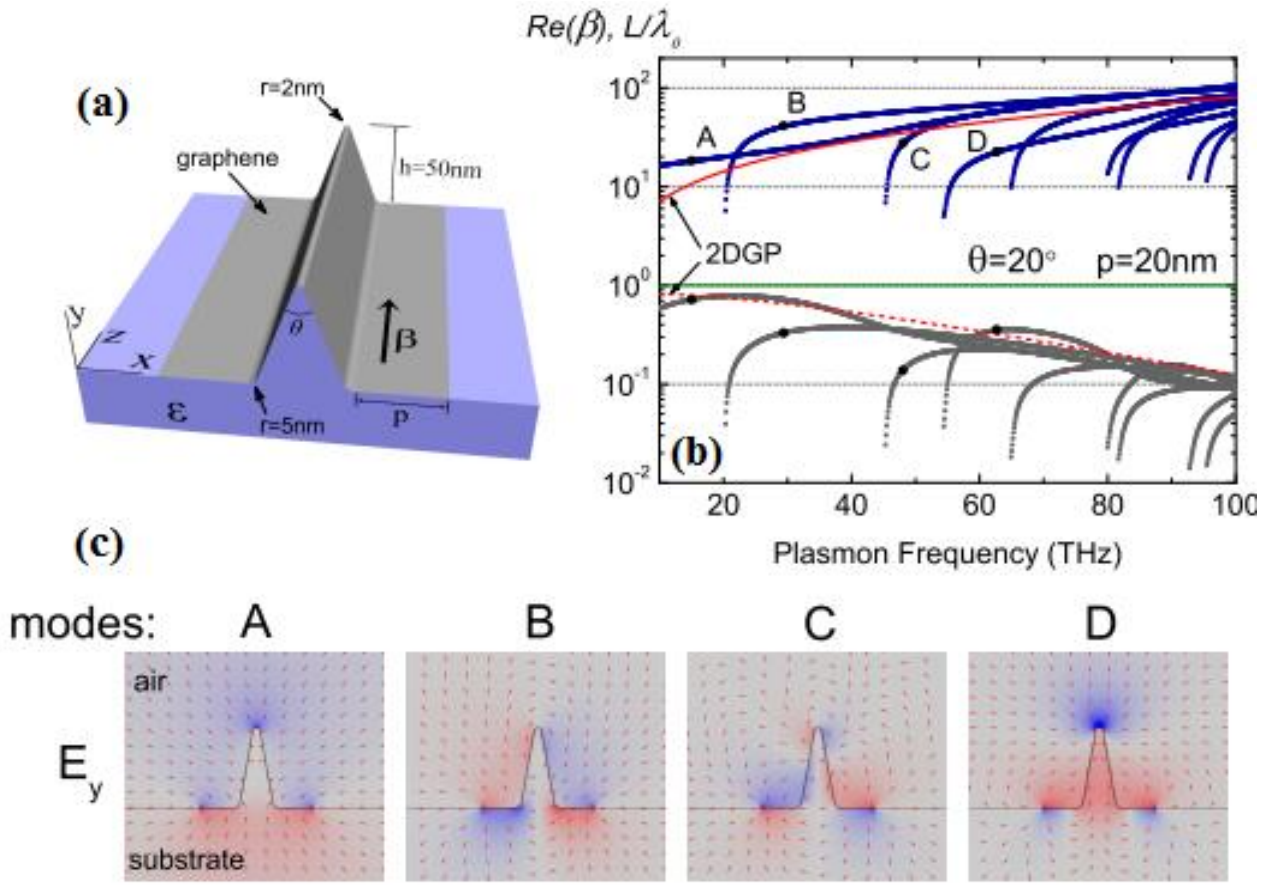

Fig. 47. (a) the schematic of wedge graphene waveguide, (b) dispersion diagram, (c) the field distribution $E_y$ for A-D modes, the geometrical parameters are $\theta = 20^0$, $p = 20\,nm$ [137]

In [138], a graphene plasmonic waveguide based on a dielectric wedge is introduced in 2013 and complete paper was published in [139]. The schematic of the proposed structure is shown in fig. 48 (a), where the graphene sheet is placed at top of the dielectric wedge [139].The mentioned structure has good confinement and low loss and can support single mode through the waveguide. The authors first exhibit the modal properties and study on results and then investigate the cross talk between two adjacent waveguides [139]. A novel hybrid plasmonic V-shaped groove is proposed by Xiaosai Wang et al in [140]. The authors utilize FEM to simulate the full structure and investigate the hybrid SPPs existed in the structure. The proposed waveguide can be used in high performance and tunable devices in infrared region [140]. A new investigation of triangular wedges and grooves are reported in [141], Similar to [137]. In this article, the authors tried to find the green's functions of structure by starting the Poisson's equation in cylindrical coordinates and employing an orthogonal polynomials expansion technique [141]. The details of calculations are given the paper [141].

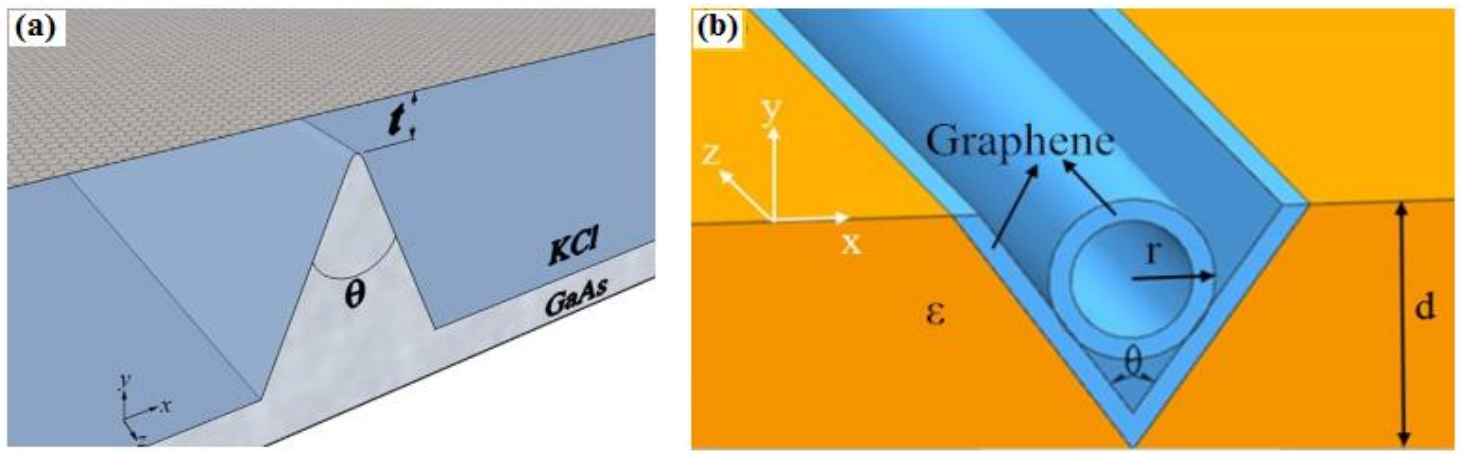

Fig. 48. (a) The geometry of the graphene sheet placed on top of the dielectric wedge [139], (b) Hybrid plasmonic V-shape groove proposed in [140]

## 3.2 Cylindrical Graphene Waveguides with Anisotropic Graphene sheet

In this section, the cylindrical graphene waveguides with anisotropic graphene sheet will be considered where the conductivity of graphene is defined as relation (5). Only two articles exist for these structures in the literature [142, 143]. Dimitrios L. Sounas and Christophe Caloz proposed a network model in [142] for an anisotropic graphene sheet in cylindrical waveguide where is located in perpendicular to cross section. The reflectance and transmission relations are obtained and depicted for a circular waveguide with radius of 8 mm [142].

In [143], anisotropic graphene based cylindrical waveguides are analyzed theoretically and their applications as reconfigurable antennas are investigated. The three graphene based cylindrical structures shown in fig. 50 are theoretically analyzed and dispersion relations are derived for them [218]. The authors utilize the structure of fig. 50(d) to design a plasmonic dipole antenna where its schematic and equivalent circuit are illustrated in fig. 51 [143]. This paper is one of the applied article which utilizes the graphene waveguide to design other devices such reconfigurable antenna.

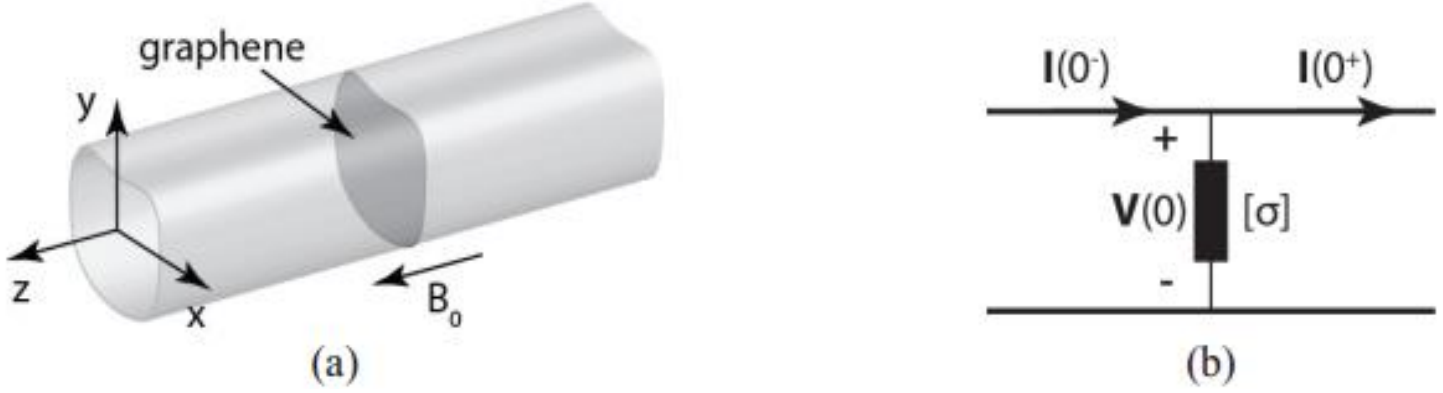

Fig. 49. (a) Cylindrical waveguide with anisotropic graphene sheet, (b) network model [142]

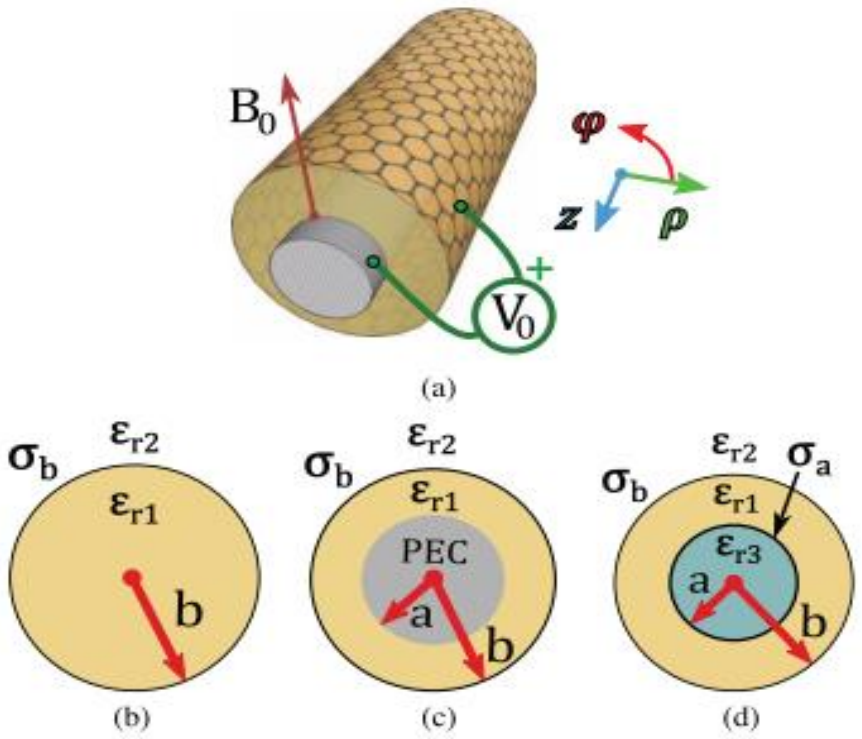

Fig. 50. (a) General graphene-based cylindrical waveguides with anisotropic graphene, cross section of three types of anisotropic graphene based cylindrical waveguides: (b) mono layer graphene cylindrical waveguide, (c) the graphene cylindrical waveguide with metallic core, (d) the graphene cylindrical waveguide with double graphene layer [143]

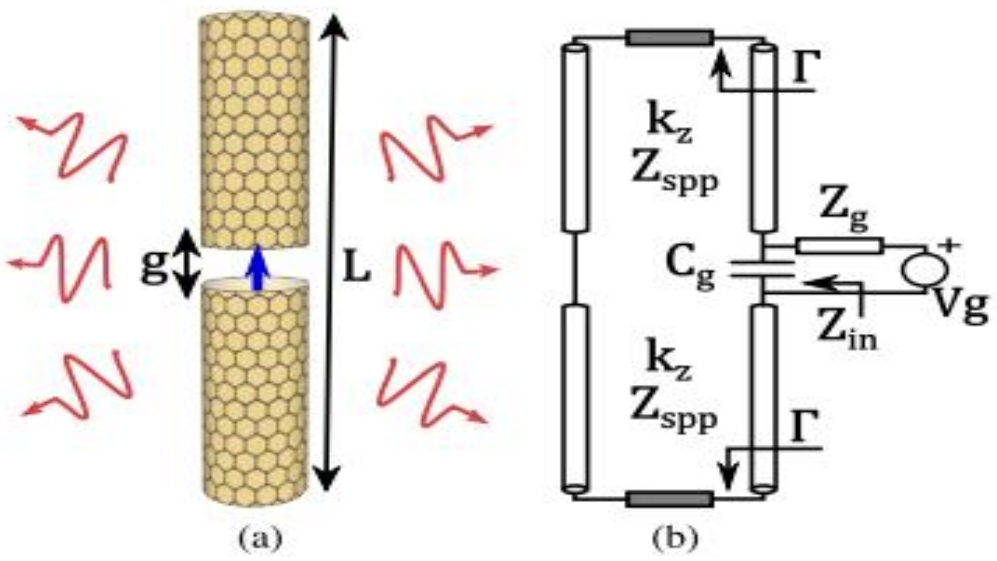

Fig. 51. (a) Reconfigurable diploe antenna based on structure of fig. 50 (d) , (b) Equivalent circuit [143]

## 4. CONCLUSION

In this paper, a historical review of plasmonic graphene waveguides is presented. The graphene based waveguides can be divided into two main categorizes: 1- rectangular and 2- cylindrical waveguides. In both categorizes, graphene can be only electrically biased or magnetically and electrically biased simultaneously where the conductivity of graphene becomes a tensor in second case. Therefore, we have four main categorizes for plasmonic graphene waveguides which are studied and discussed more precisely in this paper.